 \def\draftversion{false}

\RequirePackage{ifthen}
\ifthenelse{\equal{\draftversion}{true}}{
  \documentclass[galley,aps,pra,10pt,amsmath,amssymb,
    superscriptaddress,nofootinbib,longbibliography]{revtex4-2}
}{
  \documentclass[twocolumn,aps,prb,10pt,amsmath,amssymb,
    superscriptaddress,nofootinbib,longbibliography]{revtex4-2}
}

\usepackage{graphicx}
\usepackage[usenames,dvipsnames]{color} 
\usepackage{bm} 
\usepackage{bbm} 
\usepackage{soul} 
\usepackage{mathtools}
\usepackage[makeroom]{cancel}
\usepackage{xspace}
\usepackage{float}
\usepackage{multirow}
\usepackage{setspace}

\usepackage{hyperref}
\hypersetup{
    colorlinks=true,       
    linkcolor=blue,          
    citecolor=blue,       
    filecolor=magenta,      
    urlcolor=blue          
}

\def\Red#1{\textcolor{red}{#1}}

\ifthenelse{\equal{\draftversion}{true}}{
  \marginparwidth 2.7in
  \marginparsep 0.5in
  \newcounter{comm} 
  \def\commnext{\stepcounter{comm}}
  \def\commtext{{\bf\color{blue}[\arabic{comm}]}}
  \def\commmar{{\bf\color{blue}[\arabic{comm}]}}
  \def\aum#1{\commnext\marginpar{\small AU\commmar: #1}\commtext}
  \def\dsm#1{\commnext\marginpar{\small DS\commmar: #1}\commtext}
  \def\ytm#1{\commnext\marginpar{\small YT\commmar: #1}\commtext}
  \def\spm#1{\commnext\marginpar{\small SP\commmar: #1}\commtext}
  \def\srm#1{\commnext\marginpar{\small SR\commmar: #1}\commtext}
  \def\krm#1{\commnext\marginpar{\small KR\commmar: #1}\commtext}

  \newcommand{\eqlab}[1]{\Red{\hbox{\small\;\;[#1]}}\label{eq:#1}}
  
}{
  \def\aum#1{}
  \def\dsm#1{}
  \def\ytm#1{}
  \def\spm#1{}
  \def\srm#1{}
  \def\krm#1{}
  
  \newcommand{\eqlab}[1]{\label{eq:#1}}

}

\newcommand{\beq}{\begin{equation}}
\newcommand{\eeq}{\end{equation}}
\newcommand{\bea}{\begin{eqnarray}}
\newcommand{\eea}{\end{eqnarray}}

\newcommand{\eq}[1]{Eq.~(\ref{eq:#1})}

\renewcommand{\k}{{\bf k}}

\newcommand{\msk}{\hspace{0.8pt}}
\begin{document}

\title{\texorpdfstring{G-type Antiferromagnetic 
BiFeO$_3$ is a Multiferroic $g$-wave Altermagnet\\
}{BFO-title}}

\author{Andrea Urru}
\thanks{These authors contributed equally to the work.}
\affiliation{Department of Physics $\&$ Astronomy, Center for Materials Theory, Rutgers University, 
Piscataway, New Jersey 08854, United States}%

\author{Daniel Seleznev}
\thanks{These authors contributed equally to the work.}
\affiliation{Department of Physics $\&$ Astronomy, Center for Materials Theory, Rutgers University, 
Piscataway, New Jersey 08854, United States}%

\author{Yujia Teng}
\affiliation{Department of Physics $\&$ Astronomy, Center for Materials Theory, Rutgers University, 
Piscataway, New Jersey 08854, United States}%

\author{Se Young Park}
\affiliation{Department of Physics and Origin of Matter and Evolution of Galaxies (OMEG) Institute, Soongsil University, Seoul 06978,  Korea}

\author{Sebastian E. Reyes-Lillo}
\affiliation{Departamento de F\'isica y Astronom\'ia, Universidad Andres Bello, Santiago 837-0136, Chile}

\author{Karin M. Rabe}
\affiliation{Department of Physics $\&$ Astronomy, Center for Materials Theory, Rutgers University, 
Piscataway, New Jersey 08854, United States}

\begin{abstract}

G-type antiferromagnetic BiFeO$_3$ is shown to be an altermagnet. We present the band structure using an unconventional scheme designed to highlight the distinctive spin splitting which is characteristic of altermagnets. We define and show plots of the spin-splitting function in reciprocal space. We show that the nodal surfaces of the spin-splitting function that follow from symmetry can be classified into two types, which we call symmetry-enforced and continuity-enforced. We describe the spin-splitting function with a simple parametrization in a basis of symmetry-adapted plane waves. Using group-theory analysis based on irreducible representations of the crystallographic Laue group, we confirm that the altermagnetism of G-type BiFeO$_3$ is $g$-wave and present a complete classification table for the general three-dimensional case. Finally, we discuss the effect of ferroelectric switching on the altermagnetic order, and identify three classes of ferroelectric altermagnets.

\end{abstract}

\maketitle

\section{Introduction}

A newly identified type of magnetic order, known as altermagnetism, has recently received significant attention \cite{smejkal-prx22,smejkal-prx22-2}. Altermagnets are collinear magnetic systems that, in the limit that spin-orbit coupling (SOC) is taken to zero, are characterized by spin splitting of the electronic bands despite antiferromagnetic ordering and zero net magnetization \cite{hayami-jpsj19,hayami-prb20,yuan-prb20,smejkal-sciadv20,mazin-pnas21,yuan-prm21,yuan-advmater23,yuan-natcomm23,krempasky-nat24}.  The zero net magnetization of altermagnets is ensured by the alternation of spin splitting across the Brillouin zone (BZ), while the size of the spin splitting can be large. The spin splitting of bands in altermagnets makes them potential replacements for ferromagnets (FMs) \cite{mazin-prx22} in device engineering, particularly spintronics applications, with the advantage that, unlike FMs, they are not limited by the presence of stray magnetic fields \cite{gonzalez-prl21,smejkal-prx22-3,liu-prb24,watanabe-prb24,yang-arxiv25,sharma-arxiv25,devita-arxiv25,gunnink-arxiv25,sheng-natphys25,noh-arxiv25,leiviska-arxiv25,samanta-nanolett25,golub-arxiv25,feng-arxiv25,zhang-arxiv25,sourounis-prb25,duan-prl25}.

The symmetry of a collinear magnetically ordered crystal without SOC is fully specified by its ``magnetic space group (MSG) without SOC'' \cite{yuan-prb20,yuan-prm21,turek-prb22,yuan-natcomm23,Yuan-prl24}. For collinear magnetic ordering without SOC, the symmetry group of the Hamiltonian is expanded to the corresponding spin space group, with rotation in spin space independent of the rotation (proper or improper) in real space. Altermagnetic spin splitting is contingent on the lack of two specific symmetries in the spin space group: $\mathcal{P} \mathcal{T}$ symmetry, where $\mathcal{P}$ is spatial inversion and $\mathcal{T}$ is time reversal, and $U \textbf{t}$, where $U$ is rotation in spin space by $\pi$ around an axis perpendicular to the collinear axis, and $\textbf{t}$ is a fractional translation. Conversely, the zero spin splitting across the BZ with no SOC in the majority of familiar antiferromagnets (AFMs), including Cr$_2$O$_3$ and the G-type AFMs rocksalt-structure NiO and MnO, is the result of the presence of one or both of those symmetries in the spin space groups of those systems.

The main reason that altermagnetism was not recognized earlier as a separate type of magnetic order is that in altermagnets, the spin splitting is often zero by symmetry on high-symmetry lines and planes in the BZ. Thus, in conventional bandstructure plots, which follow a path around the BZ formed by high-symmetry lines \cite{hinuma-cms17}, it can be expected that no spin splitting will be visible; this is particularly likely for higher-symmetry magnetic systems, which tend to be favored for study. The reasons for this convention are historical: in the early days of bandstructure calculation, symmetry was heavily used to make computations tractable. In addition, early work focused on the analysis of critical points in the density of states, focusing interest on band extrema that usually (though not always) are at high-symmetry points. However, if the goal is a representative sampling of the bands in the BZ, this is not the correct approach: the high-symmetry lines are the least representative, with the $\Gamma$ point being the very least representative of all. The time is right to overturn this convention, and below, we introduce a new general path in reciprocal space that mixes high-symmetry lines and general lines, in a way that makes clear the presence of spin splitting and the alternation of its sign across the BZ, a key signature of altermagnets. 

\begin{figure*}[t!] 
\includegraphics[width=0.75\textwidth]{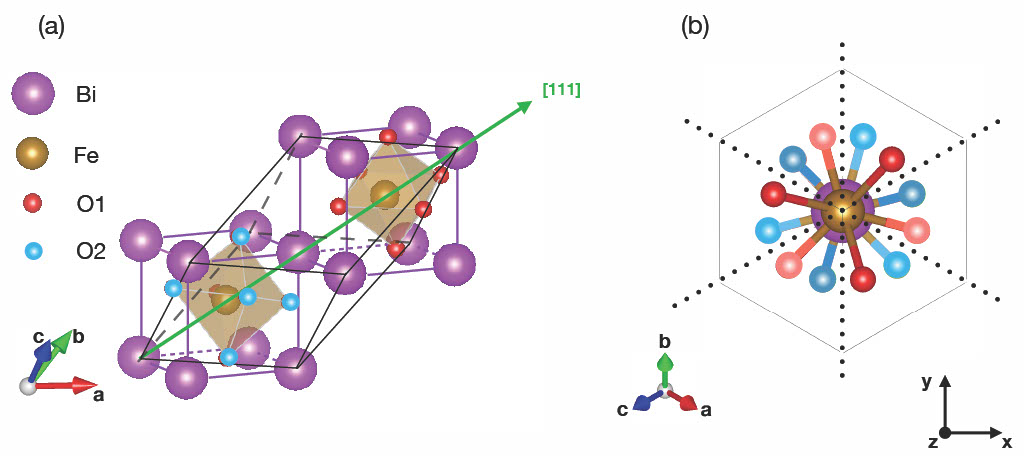}
\caption{\label{fig:structure} (a) Three-dimensional view of the rhombohedral R3$c$ structure of BFO, highlighting its relation to the ideal cubic perovskite structure. The 10-atom rhombohedral unit cell is shown by black lines. Perovskite-like constitutive blocks are indicated with solid purple lines and FeO$_6$ octahedra are highlighted in brown. Oxygen atoms belonging to the two different octahedra are shown in red and blue. The three-fold axis along the [111] pseudo-cubic direction is identified by a solid green arrow. (b) View down the [111] direction. 
The lighter and darker colors for the oxygen atoms indicate atoms nearer to and farther from the viewer, respectively. The $c$ glide planes are highlighted with dotted black lines. The Cartesian reference frame used in the text is also shown.}
\end{figure*}

A case in point illustrating the failure of conventional bandstructure paths for the identification of altermagnetism is the multiferroic G-type AFM BiFeO$_3$ (BFO). Despite a vast literature reporting studies of BFO since 2003 \cite{wang-science03, neaton-prb05, ederer-prb05,burns-adm20}, the altermagnetic spin splitting in BFO went unnoticed until the recent surge of interest in altermagnetism \cite{bernardini-jap25,smejkal-arxiv24,dong-prb25,dong-arxiv25} because the spin splitting is zero on the high-symmetry lines on the conventional bandstructure path \cite{neaton-prb05}. However, as we will see, the altermagnetic spin splitting is easily recognized in our proposed bandstructure plotting convention. 

Further investigation of the altermagnetism in G-type AFM BFO brings many interesting issues to light. We define a spin-splitting function on the BZ and present maps of the spin splitting on planar cuts of the BZ. We identify the nodal surfaces of the spin-splitting function and discuss how this map shows the $g$-wave character of altermagnetism in BFO, with the caveat that looking at a single plane can lead to confusion. We show how to unequivocally identify the ($d$,$g$,$i$)-wave character from the spin Laue group alone and provide a complete table. We further analyze the spin-splitting function, which is periodic in reciprocal space, via a Fourier analysis using symmetry-adapted linear combinations of plane waves, generalizing the seminal works of Refs.~\cite{chadi-prb73_1,chadi-prb73_2,baldereschi-prb73,joannopolous-jpc73,monkhorst-prb76}. 

A key driver of interest in BFO has been its multiferroic character. BFO combines ferroelectricity and weak ferromagnetism at room temperature, addressing one of the major challenges of the physics of multiferroics, which is the contraindication between ferroelectricity and ferromagnetism \cite{spaldin-jphyschemb00,fiebig-natrevmat16,spaldin-natmat2019}. In contrast, ferroelectricity and altermagnetism are quite compatible \cite{smejkal-arxiv24,sun-am25,wang-arxiv25}. For one thing, the atomic arrangement of ferroelectrics breaks $\mathcal{P}$, and thus $\mathcal{P}\mathcal{T}$ is always absent from the spin space group, so that the symmetry criterion for altermagnetism is simplified to the absence of $U \textbf{t}$. In addition, the insulating character needed for ferroelectricity is naturally compatible with the AFM ordering in altermagnets. Motivated by recent work on the control of spin-splitting in hybrid improper multiferroics \cite{benedek-prl11} through rotational modes of oxygen octahedra surrounding the magnetic atoms \cite{gu-prl25,smejkal-arxiv24}, we explore how the spin splitting in BFO is affected by the structural distortions leading from the high-symmetry cubic phase of BFO to its low-symmetry rhombohedral phase. This investigation of BFO thus offers promise for rich physics in this new class of multiferroics. 

The remainder of the manuscript is organized as follows. In Sec.~\ref{sec:symm}, we analyze the symmetry of BFO and explain how it allows for altermagnetism. In Sec.~\ref{sec:bands}, we introduce the aforementioned generalized path in reciprocal space and present the band structure of BFO in absence of SOC (Sec.~\ref{sec:bands-no-soc}) showing its sizeable spin splitting along general lines; we also show the band structure in the presence of SOC (Sec.~\ref{sec:bands-soc}), which highlights how SOC breaks spin space group symmetries and further splits the altermagnetic bands.  In Sec.~\ref{sec:heatmaps}, we define the spin-splitting function as a quantitative description of the momentum-dependent spin splitting across the BZ. We analyze the symmetries and nodal surfaces of the spin-splitting function and introduce a symmetry-adapted plane-wave expansion that leads to a simple parametrization; we then apply this to analyze the altermagnetic spin splitting in BFO across the BZ. In Sec.~\ref{SS-classification}, we discuss the $d$-$g$-$i$-wave classification of altermagnets using the spin-splitting function. We present a complete table with classification according to the spin Laue group and definitively classify BFO as $g$-wave. In Sec.~\ref{sec:struct-dist}, we show how the spin splitting is controlled by the structural distortion modes of the parent cubic phase of BFO. Finally, Sec.~\ref{sec:conclusions} contains a summary and conclusions. The computational details of the \textit{ab initio} calculations presented in this work are reported in the Supplementary Material (SM). 

\section{\texorpdfstring{Altermagnetism in B\lowercase{i}F\lowercase{e}O$_3$: symmetry considerations}{Symm-BFO}}
\label{sec:symm}

In this section, we describe the crystal structure and magnetic ordering of BFO, discuss its symmetries, and explain how its symmetry group allows for altermagnetism. 
BFO crystallizes in a R$3c$ rhombohedral structure, shown in Fig.~\ref{fig:structure}. 
This structure can be obtained from the P$m\bar{3}m$ ideal cubic perovskite structure by an $R$-point pattern ($R_4^+$) of alternating oxygen octahedron rotations around [111] and a polar distortion ($\Gamma_4^-$) along [111]. The symmetries of this atomic arrangement include a three-fold rotation axis along the [111] pseudo-cubic direction (Fig.~\ref{fig:structure}(a)) and three $c$-glide planes (Fig.~\ref{fig:structure}(b)) intersecting at the three-fold axis. 

The experimentally reported ground-state magnetic ordering of BFO is a spin cycloid with a period of $\approx 62$ nm~\cite{Sosnowska-jphysc82}. In thin films, this long-period cycloid is suppressed and the magnetic ordering is reported as AFM G-type, with two Fe atoms in the rhombohedral unit cell having opposite spins \cite{burns-adm20}. First-principles calculations confirm G-type AFM ordering \cite{neaton-prb05}, with weak ferromagnetism arising from spin canting when SOC is taken into account \cite{ederer-prb05}. In addition, experimental measurements~\cite{fu-apl14} and first-principles calculations~\cite{dupe-prb10} suggest that epitaxial strain further favors the G-type AFM order. 

The ``MSG without SOC'' of G-type BFO is R$3c'$ (type III, No.\ 161.71). The R$3c$ rhombohedral crystal structure is non-centrosymmetric, and thus the spin space group of G-type BFO lacks $\mathcal{PT}$. Moreover, because of the alternating oxygen octahedron rotations, the two Fe sublattices are not connected by a pure translation, and thus the spin space group of G-type BFO lacks $U\textbf{t}$. Instead, the up and down Fe sublattices are mapped into each other by the $c$ glide mirror planes, implying zero net magnetization. Thus, G-type BFO satisfies the symmetry requirements to be an altermagnet. For completeness, we mention here that the spin point group~\cite{brinkman-prsl66,litvin-physica74,andreev-jetp76,litvin-actacrysta77,andreev-spu80} is ${}^13^2m$, which implies $g$-wave type altermagnetism~\cite{smejkal-prx22}; we will return to this point below.
 
The symmetry analysis is modified when SOC is included. In this case, the symmetry of the Hamiltonian is specified by its MSG, with rotations in spin space locked to rotations in real space. Consequently, the MSG depends on the real-space orientation of the magnetic dipole moments. For G-type BFO with Fe magnetic moments along the three-fold axis, the MSG is R$3c.1$ (type I, No.\ 161.69). If G-type ordering is established with the magnetic moments in the plane perpendicular to the crystallographic three-fold axis, the three-fold symmetry is broken and there is at most a single glide plane. If the G-type magnetic moments are perpendicular to the glide plane, the MSG is C$c'$ (type III, No.\ 9.39), which allows for nonzero FM moment in the glide plane. On the other hand, if the G-type magnetic moments are in the glide plane, the MSG is C$c.1$ (type I, No.\ 9.37), which allows for nonzero FM moment along the direction perpendicular to the glide plane. In both cases, the FM moment is a weak effect coming from spin canting induced by SOC, while the primary order remains AFM. It has been reported from first principles calculations that C$c'$ and C$c.1$ are very similar in energy, with both lower in energy than R$3c.1$ \cite{ederer-prb05}. In the next section, we assume C$c'$ for our consideration of SOC effects, taking the AFM component of the spins to be along the Cartesian $x$ direction defined in Fig.~\ref{fig:structure}(b). 

\begin{figure}[t] 
\includegraphics[width=0.35\textwidth]{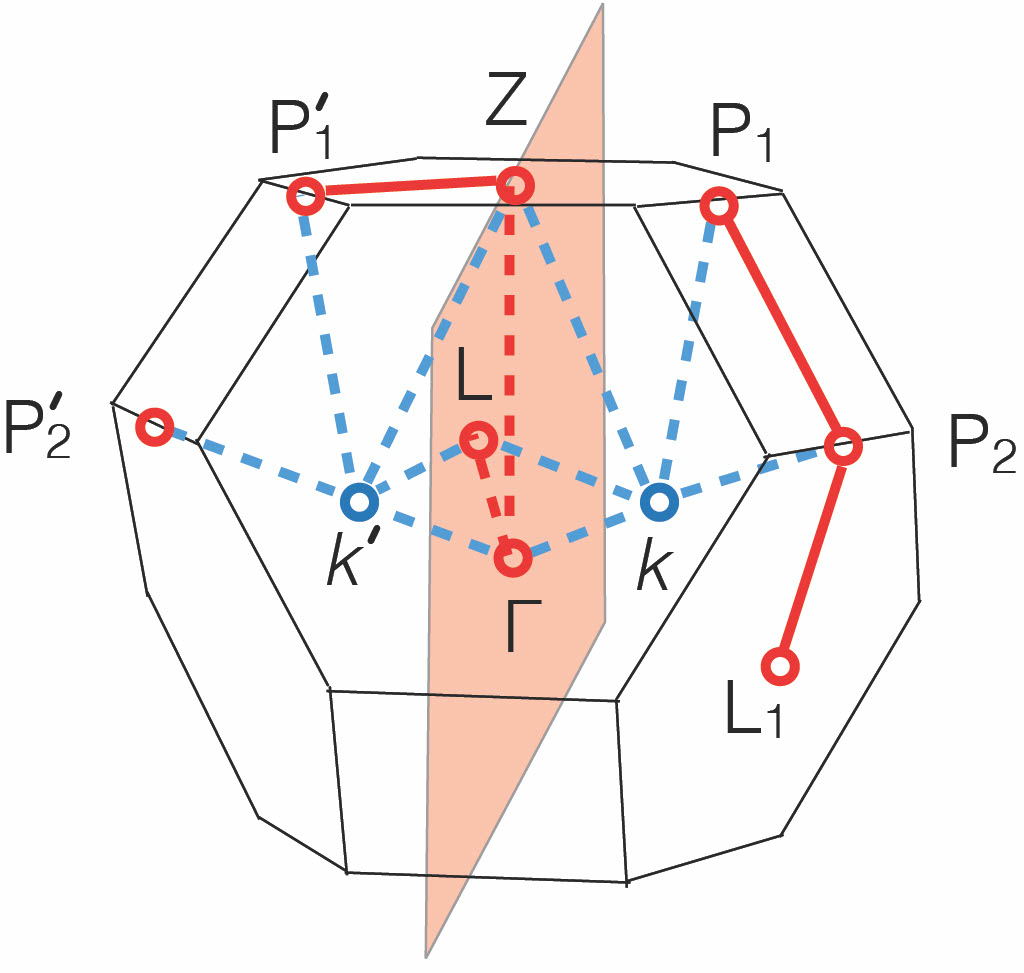}
\caption{\label{fig:BZ} Brillouin zone of the rhombohedral lattice. High-symmetry points are identified by red open circles. Red dashed lines indicate high-symmetry lines residing in the interior of the Brillouin zone, while solid red lines indicate those residing on its boundary. Blue dashed lines indicate general lines connecting high-symmetry points with the general points $k$ and $k'$, identified with blue open circles. Primed and un-primed points are related by the mirror plane shown in pink.}
\end{figure}

\section{Altermagnetic spin splitting in the electronic band structure}
\label{sec:bands}

In this section, we present the electronic band structure of BFO computed from first principles. We highlight the presence of an altermagnetic spin splitting in absence of SOC, and then discuss how SOC further splits the altermagnetic bands.  

\subsection{Without spin-orbit coupling}
\label{sec:bands-no-soc}

\begin{figure*}[t!] 
\includegraphics[width=0.75\textwidth]{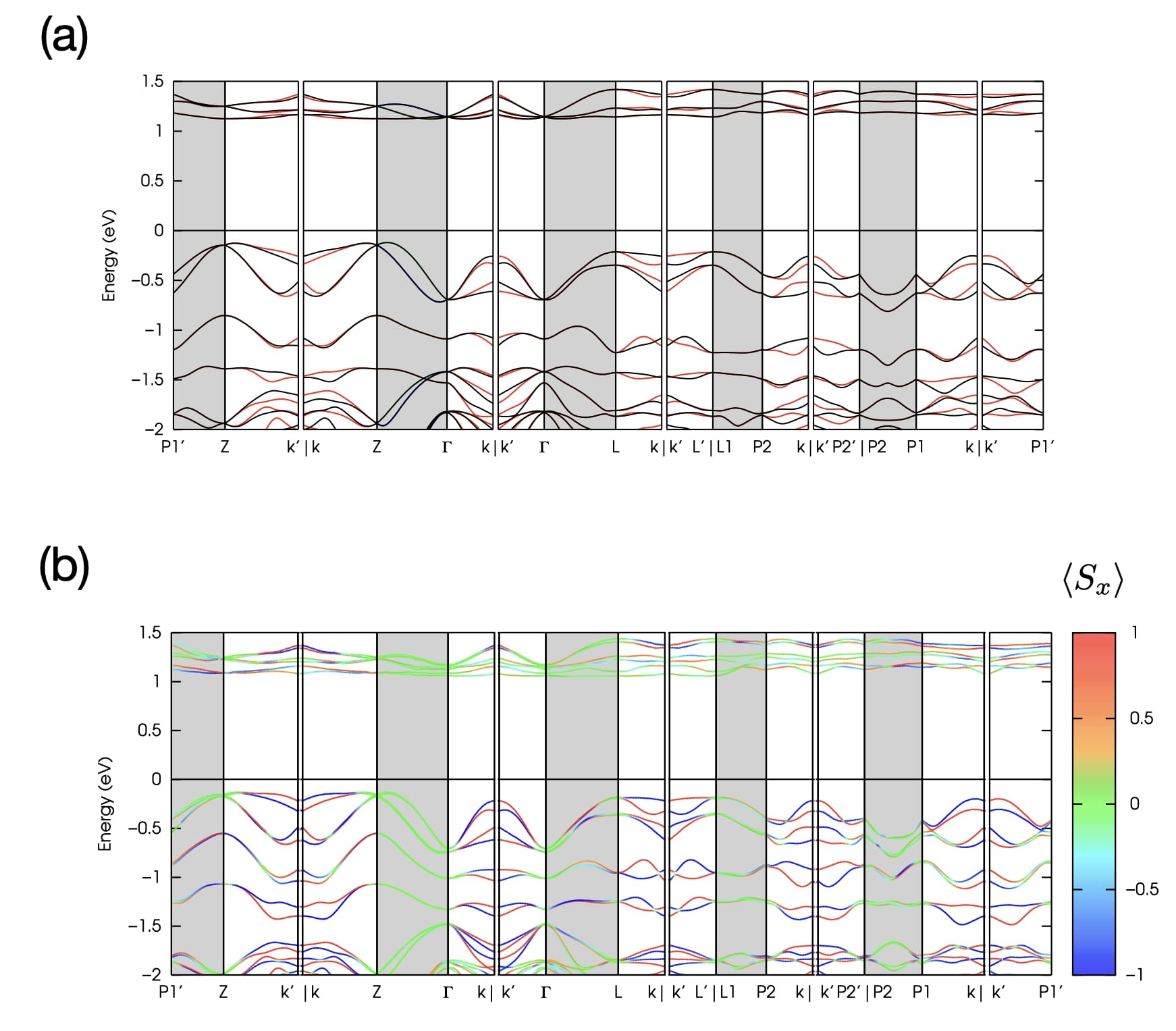}
\caption{\label{fig:bands} (a) Band structure of G-type BFO computed without SOC shown along a generalized path in reciprocal space, including high-symmetry lines (grey panels) and general lines (white panels). (b) Band structure of G-type BFO computed, with SOC, along the same generalized path in reciprocal space. Bands are colored according to the expectation value of the spin along the $x$ direction.}
\end{figure*}

In the absence of SOC, the spin point group of G-type BFO is ${}^13^2m$. Symmetry considerations imply that the spin splitting for all high-symmetry $\textbf{k}$-points in the BZ (Fig.~\ref{fig:BZ}) is zero. Specifically, all these $\textbf{k}$-points, whose coordinates we list in the SM, lie on at least one mirror plane, with the mirror operation combined with spin reversal ($^2m$ in the spin point group symbol). This spin point group symmetry originates from the ``MSG without SOC" glide planes $c'$, which combine a mirror operation with the fractional translation $(1/2,1/2,1/2)$, linking the up and down Fe sublattices. Since this symmetry connects the two sublattices in real space, it transforms spin-up bands into spin-down bands in reciprocal space, thus protecting spin degeneracy along the high-symmetry lines. The alternation of spin splitting around the BZ also follows from the transformation of spin-up bands to spin-down bands by the mirror planes. Each general point \textbf{k} has a counterpart $\mathbf{k}'$ related by a mirror plane as shown in Fig.~\ref{fig:BZ}. Since the mirror plane maps spin-up states into spin-down states, the spin splitting at $\mathbf{k}'$ is reversed with respect to that at \textbf{k}, a key feature of altermagnetism.   

Since the conventional path for R$3c$ includes only high-symmetry lines, the spin splitting is not apparent in a conventional bandstructure plot \cite{neaton-prb05,hinuma-cms17}. Here, we propose and demonstrate an alternative scheme for plotting the band structure. We follow a path in the BZ that centers on a general \textbf{k}-point \textit{k}, rather than $\Gamma$. \textit{k} is chosen to be inside the irreducible wedge, far from the high-symmetry planes and lines. For G-type BFO, we use $k = (-0.379, 0.070, 0.101)$, in units of $2 \pi / a$, where $a = 5.668$ \AA\ is the lattice constant of the rhombohedral cell. The path includes lines that connect \textit{k} to high-symmetry points ($\Gamma$, $Z$, $L$, $L_1$, $P_1$, $P_2$) and also includes segments that connect high-symmetry points along high-symmetry lines as the path traverses the BZ.   
For systems with altermagnetic spin space groups, we identify a point $k'$ related to the general point $k$ by a spin point group element ${}^2R$ that combines a proper or improper rotation $R$ with a spin reversal (for G-type BFO this is a mirror plane $^2m$), and pair each line connecting $k$ to a high-symmetry point with the image of the line under $^2R$. The purpose of this pairing is to exhibit the alternation of spin splitting across the BZ and the resulting zero net magnetization.

In Fig.~\ref{fig:bands}(a), we present the computed band structure of G-AFM BFO without SOC, using the alternative scheme just described.
A spin splitting of magnitude up to $\approx 0.2$ eV is seen on general lines in Fig.~\ref{fig:bands}(a), with alternation across the BZ, while the bands are spin-degenerate along high-symmetry lines, as expected.
This scheme thus allows ready identification of altermagnets and, by sampling general \textbf{k}-points as well as high-symmetry lines, has the additional advantage of being more representative of the band structure than the conventional scheme.

\subsection{With spin-orbit coupling}
\label{sec:bands-soc}

As discussed in Sec.~\ref{sec:symm}, we report the computed band structure for G-type BFO with SOC included in magnetic space group C$c'$, with spin components along the $x$ direction (see Fig.~\ref{fig:BZ}) showing G-type AFM ordering \cite{ederer-prb05}. We performed fully relativistic density-functional theory (DFT) calculations with spinor wave functions. SOC-induced spin canting gives weak FM components along $y$ and along $z$. The computation is done assuming the same R$3c$ structure as for zero SOC; we have checked relaxation of the structure with the lower symmetry C$c'$ and found it leads to negligible changes. Our calculations show that the net magnetic dipole per unit cell along the $y$ direction is $\approx 0.09 \, \mu_{\text{B}}$, in agreement with the value of $\approx 0.1 \, \mu_{\text{B}}$ reported in Ref.~\cite{ederer-prb05}, while the net $z$ component of the magnetic dipole, although allowed by symmetry, is negligible ($< 0.001 \, \mu_{\text{B}}$) \cite{ederer-prb05}.

In Fig.~\ref{fig:bands}(b), we show the band structure computed for C$c'$ G-type BFO including SOC on the same BZ path as in Fig.~\ref{fig:bands}(a). The bands are colored according to the expectation value of the spin along $x$, the direction of the AFM component of magnetic dipoles. The spin character of the bands along the general lines is not substantially perturbed by SOC; the bands have mostly spin-up and spin-down character along the $x$ axis, and retain to a large extent the altermagnetic features that we highlighted previously in absence of SOC. The spin degeneracy seen along the high-symmetry lines in Fig.~\ref{fig:bands}(a) is broken and a small spin splitting is visible, except at the $Z$ point, as expected from symmetry considerations \cite{Aroyo-bcc2011, Aroyo-ACA2006, Aroyo-zkri2006, Xu-nat20, Elcoro-natcomm21}. Along general lines, the SOC-induced splitting appears on top of the non-relativistic one that we already discussed previously. Overall, the SOC-induced spin splitting has an average magnitude approximately five times smaller than the non-relativistic splitting on the valence band manifold, whereas the two splittings are of comparable size on the conduction bands. 

\section{Spin splitting in reciprocal space}
\label{sec:heatmaps}
With Sec.~\ref{sec:bands} having established the presence of altermagnetic spin splitting in the electronic band structure of BFO, this section is devoted to analyzing it in a detailed and quantitative manner, focusing on its appearance in the entire BZ rather than selected general or high-symmetry lines in $\textbf{k}$-space. To do so, we first define the $\textbf{k}$-dependent spin-splitting function and explore its general symmetry properties in altermagnets. Using these general symmetry properties, we subsequently explore the spin-splitting function in BFO and analyze its nodal surfaces. 
Finally, we construct a basis set of symmetry-adapted plane waves for a general description and simple parametrization of the spin splitting functions, and show the results for BFO. 

\subsection{General symmetry properties of altermagnetic spin splitting}\label{sec:SS-symmetries}
We consider a composite group of $N$ pairs of spin-split bands that are well separated from all other bands. This could be, for example, the entire valence manifold of an insulator, or some other isolated subset of bands. We define the $\textbf{k}$-dependent spin splitting of this set of bands as \begin{equation}\Delta(\textbf{k})=\frac{1}{N}\sum^{N}_{n=1}\left[\epsilon_{n,\uparrow}(\textbf{k})-\epsilon_{n,\downarrow}(\textbf{k})\right],\label{eq:SS-splitting}\end{equation} where $\epsilon_{n,\uparrow}(\textbf{k})$ and $\epsilon_{n,\downarrow}(\textbf{k})$ are the $n$-th spin-up and spin-down band energies, respectively. 

While the symmetry of an altermagnetic system as a whole is described by a spin space group, the symmetry properties of associated functions defined on the BZ, such as $\Delta(\textbf{k})$, are determined by the corresponding spin point group \cite{brinkman-prsl66,litvin-physica74,andreev-jetp76,litvin-actacrysta77,andreev-spu80,liu-prx22,watanabe-prb24,xiao-prx24}. Elements of the spin point and spin space groups are composed of pairs of transformations $[S_i||S_j]$, where the transformation to the left of the double vertical bar acts solely on spin space, while the one to the right acts on real space. Operations involving time reversal $\mathcal{T}$ will be denoted as $\mathcal{T}[S_i||S_j]$, since $\mathcal{T}$ always acts both on spin space and real space. Upon inclusion of SOC, spin space and real space become coupled, and the spin space groups reduce to MSGs containing elements of the form $[S_i||S_i]$ or $\mathcal{T}[S_i||S_i]$. In what follows, we focus on the limit of vanishing SOC.

Collinear altermagnets feature several general spin space symmetries that impose constraints on the symmetry properties of the spin-splitting function. First, as the magnetic sublattices in altermagnets are related to each other solely by proper or improper spatial rotations (possibly combined with fractional translations), the spin-splitting function is invariant under operations of the form $[C_2||S]$. Here, $C_2$ is a global $\pi$ rotation in spin space about an axis perpendicular to the collinear axis \cite{litvin-physica74,litvin-actacrysta77}, while $S$ is the rotation relating the up and down magnetic sublattices. Second, altermagnets are invariant under $\mathcal{T}[C_2||E]$ (where $E$ is the identity acting in real space) as are indeed all collinear AFMs. Finally, the combination of the two preceding symmetries -- $\mathcal{T}[E||S]$, where $E$ is now the identity acting in spin space -- must also be a symmetry under group closure.

Under these symmetries, the spin splitting obeys 
\begin{align}
\Delta(\textbf{k})&=[C_2||S]\Delta(\textbf{k})=-\Delta(S\textbf{k}),\label{eq:SpinRot}\\
\Delta(\textbf{k})&=\mathcal{T}[C_2||E]\Delta(\textbf{k})=\Delta(-\textbf{k}),\label{eq:SpinTime}\\
\Delta(\textbf{k})&=\mathcal{T}[E||S]\Delta(\textbf{k})=-\Delta(-S\textbf{k}).\label{eq:RotTime}
\end{align}
Equations (\ref{eq:SpinRot}) and (\ref{eq:RotTime}) indicate that $\Delta(\mathbf{k})$ alternates in sign throughout the BZ. The two symmetries additionally ensure the existence of regions in reciprocal space where the spin splitting vanishes. Namely, the sets of points where $S\textbf{k}\equiv\textbf{k}$ and $S\textbf{k}\equiv-\textbf{k}$ (where equivalence is up to a reciprocal lattice vector) combine to form symmetry-enforced nodal regions for $\Delta(\mathbf{k})$. We note, however, that accidental nodal features that are not enforced by symmetry can also arise.

From Eq.~(\ref{eq:SpinTime}) we observe that $\mathcal{T}[C_2||E]$ symmetry forces $\Delta(\mathbf{k})$ to be an even function of $\textbf{k}$. Notably, $\mathcal{T}[C_2||E]$ acts on $\Delta(\mathbf{k})$ in the same way as spatial inversion, $[E||\mathcal{P}]$. 
Although $\mathcal{PT}$ is broken in altermagnets, $[E||\mathcal{P}]$ can nonetheless belong to the system’s spin point group if an inversion center lies on a magnetic sublattice. 
In light of this fact, it is impossible to recognize whether or not $[E||\mathcal{P}]$ belongs to the spin point group when analyzing the symmetries of $\Delta(\textbf{k})$, as its action on the function is indistinguishable from that of $\mathcal{T}[C_2||E]$. The symmetries of $\Delta(\mathbf{k})$ are therefore characterized by the spin Laue groups, rather than by the spin point groups. A spin Laue group is formed by taking all elements of a spin point group and adjoining to it the set obtained by composing each element with $[E||\mathcal{P}]$. If $[E||\mathcal{P}]$ is already contained in the spin point group, then the spin Laue group coincides with it. On a final note, we also observe that in the context of the spin Laue group, the actions of $[C_2||S]$ and $\mathcal{T}[E||S]$ on $\Delta(\textbf{k})$ are indistinguishable from that of $\mathcal{T}[E||S\mathcal{P}]$ and $[C_2||S\mathcal{P}]$, respectively.

\subsection{Symmetries and nodal structure of altermagnetic spin splitting in BFO}\label{sec:BFO-SS-symmetries}

Having established the general symmetry properties of the altermagnetic spin-splitting function $\Delta(\textbf{k})$, we now analyze its specific symmetries and nodal structure in the context of BFO. As indicated in Secs.~\ref{sec:symm} and \ref{sec:bands-no-soc}, the symmetry group of BFO without SOC is the spin space group with parent MSG R$3c'$, with the corresponding spin Laue group being ${}^1\bar{3}^2m$. Certain symmetry elements in ${}^1\bar{3}^2m$, either individually or in combination, enforce the presence of nodal regions for the spin-splitting function. These are $[C_2||M_x]$, $[C_2||C_{2x}]$, and the three-fold rotation $[E||C_{3z}]$.

The $[C_2||M_x]$ symmetry, where $M_x$ in combination with $\textbf{t}=(1/2,1/2,1/2)$ exchanges the magnetic sublattices, enforces nodes at $\textbf{k}$-points that satisfy $M_x\textbf{k}\equiv\textbf{k}$. All points belonging to the $k_y$-$k_z$ plane respect this condition. Two additional nodal planes, related to the $k_y$-$k_z$ plane by successive three-fold rotations about the $k_z$ axis, emerge by virtue of the $[E||C_{3z}]$ symmetry. We will refer to these three nodal planes as mirror nodal planes.

$[C_2||C_{2x}]$ enforces nodes at $\textbf{k}$-points satisfying $C_{2x}\textbf{k}\equiv\textbf{k}$. All points residing on the $k_x$-axis satisfy this condition. In addition, the lines parallel to the $k_x$-axis and passing through the $Z$ point on the BZ boundary (see Fig.~\ref{fig:BZ}) satisfy this criterion. Two successive applications of $[E||C_{3z}]$ to each of the aforementioned lines result in the emergence of additional nodal lines. All of these lines shall henceforth be collectively referred to as two-fold axis nodal lines. Due to the periodicity of $\Delta(\textbf{k})$ in reciprocal space, we also note that periodic images of the two-fold axis nodal lines appear in two planes perpendicular to the $\Gamma$–$Z$ direction, located at one-third and two-thirds of the distance between $\Gamma$ and $Z$, respectively. This same periodicity causes the spin-splitting pattern in the former plane to be related by an in-plane displacement to that in the plane containing the $Z$ point, while the spin-splitting pattern in the latter plane is similarly displaced relative to the splitting in the $k_z = 0$ plane.

\begin{figure*}[t] 
\includegraphics[width=0.82\textwidth]{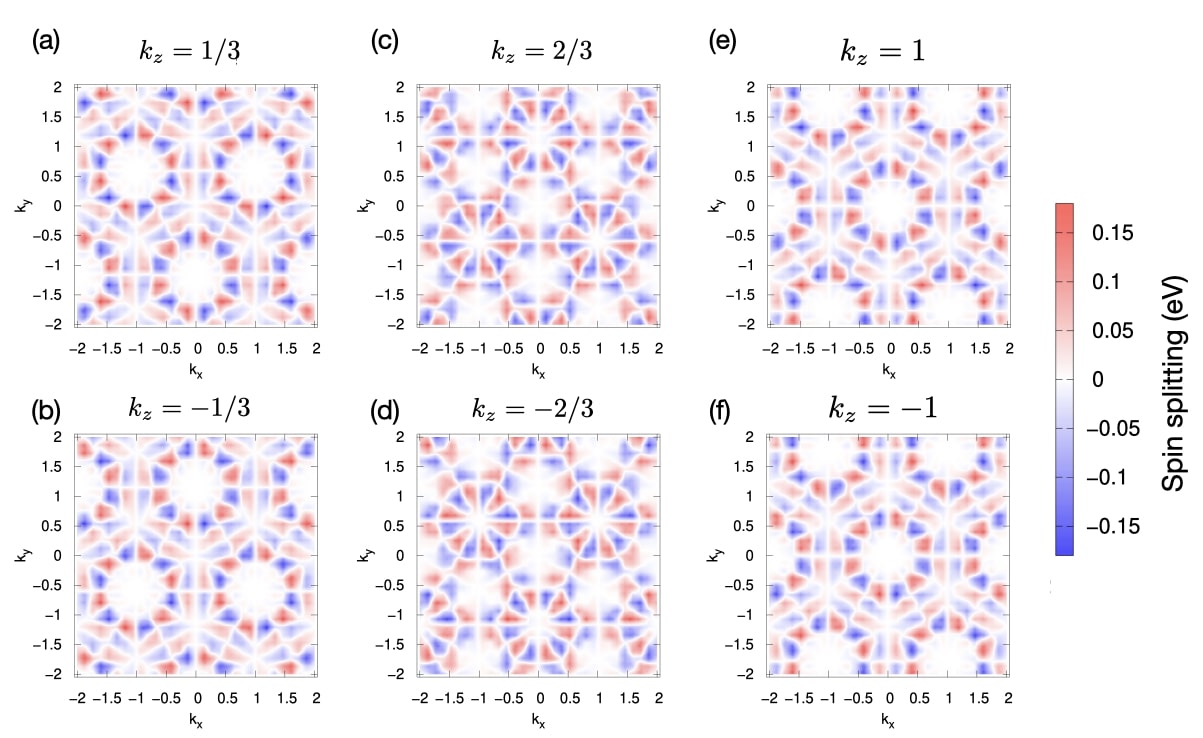}
\caption{\label{fig:heatmaps} Heatmaps showing the spin-splitting function, as defined in the main text, on portions of the BZ at constant $k_z$ values. The $k_x$ and $k_y$ coordinates are in units of $2 \pi / a$, with $a = 5.668$ \AA\ being the lattice constant of the rhombohedral cell. The $k_z$ coordinate is given as a fraction of the $\Gamma$-$Z$ distance.}
\end{figure*}

\begin{figure}[b] 
\includegraphics[width=0.5\textwidth]{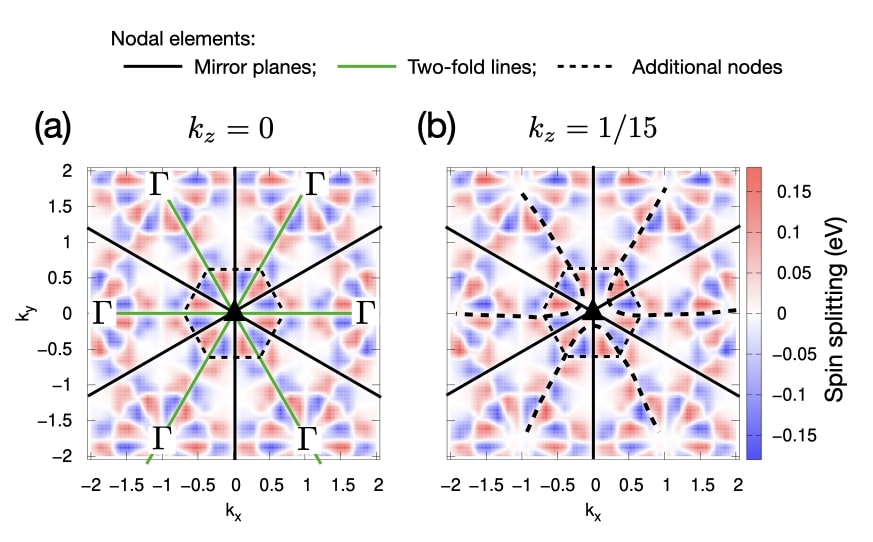}
\caption{\label{fig:nodal_highlight} Heatmap of $\Delta(\mathbf{k})$ at (a) $k_z = 0$ and (b) $k_z = 1/15$, with nodal elements highlighted. The $k_x$ and $k_y$ coordinates are in units of $2 \pi / a$, while the $k_z$ coordinate is given as a fraction of the $\Gamma$-$Z$ distance, as in Fig.~\ref{fig:heatmaps}. The three-fold axis is identified by the black triangle, whereas the mirror planes and two-fold axes, which give rise to symmetry-enforced nodal planes and two-fold nodal lines, are identified with black and green solid lines, respectively. Additional continuity-enforced nodes are highlighted with dashed black lines.}
\end{figure}

To verify the presence of these symmetries and nodes in the BFO spin-splitting function, we present in Figs.~\ref{fig:heatmaps} and \ref{fig:nodal_highlight} a series of heatmaps visualizing $\Delta(\textbf{k})$ across two-dimensional slices of the BZ parallel to the $k_x$-$k_y$ plane. The spin splitting is shown for the third pair of valence bands below the Fermi level in the BFO bandstructure, since, as seen in Fig.~\ref{fig:bands}, this pair is well separated from the rest of the valence band manifold. The band energies are computed from a non-self-consistent DFT calculation performed on a uniform $31\times31\times31$ $\Gamma$-centered $\textbf{k}$-mesh. We obtain the spin-splitting function from the values on this mesh by interpolation using the ``pm3d'' algorithm, to obtain a smooth function over the BZ.

The $k_z$ slices shown in Fig.~\ref{fig:heatmaps} are centered on points lying along the $\Gamma$-$Z$ high-symmetry line. The slices correspond to $k_z = \pm1/3$ (panels (a) and (b)), $k_z = \pm2/3$ (panels (c) and (d)), and $k_z = \pm1$ (panels (e) and (f)), where $k_z$ is given as a fraction of the $\Gamma$-$Z$ distance. All heatmaps clearly demonstrate $[E||C_{3z}]$, $[C_2||M_x]$, and $[C_2||C_{2x}]$ symmetries and their combinations. The $[E||\mathcal{P}]$ symmetry may be observed by comparing heatmaps at slices with opposing signs of $k_z$; for a given $k_z$, the splitting at $(k_x,k_y)$ is the same as that at $(-k_x,-k_y)$ in the $-k_z$ slice. The pair of slices depicted in Figs.~\ref{fig:heatmaps}(e) and (f) are furthermore identical; this is because they pass through opposing BZ boundaries and are thus separated by a reciprocal lattice vector. As explained earlier, the planes at $k_z=\pm1$ and $k_z=\pm1/3$ are also connected by reciprocal lattice vectors with components in $k_x$ and $k_y$. This results in the spin-splitting patterns in the planes in Figs.~\ref{fig:heatmaps}(a) and (e), and also Figs.~\ref{fig:heatmaps}(b) and (f), being displaced in-plane relative to one another.

The three mirror nodal planes appear as straight nodal lines passing through the centers of the heatmaps in Fig.~\ref{fig:heatmaps}. Figures \ref{fig:heatmaps}(a)-(f) also feature nodal lines that are two-fold nodal axis lines, as discussed above. Panel (a) in Fig.~\ref{fig:nodal_highlight} depicts the spin-splitting heatmap at $k_z=0$. Here, the $[E||C_{3z}]$, $[C_2||M_x]$, and $[C_2||C_{2x}]$ symmetries, whose presence is evident from the heatmap, are explicitly highlighted.
 We further note that the heatmaps in Figs.~\ref{fig:heatmaps}(c) and (d) exhibit spin splitting that is displaced in-plane relative to that in Fig.~\ref{fig:nodal_highlight}(a), as the former planes are related to the latter by reciprocal lattice vectors with in-plane components.

\begin{figure}[t] 
\includegraphics[width=0.4\textwidth]{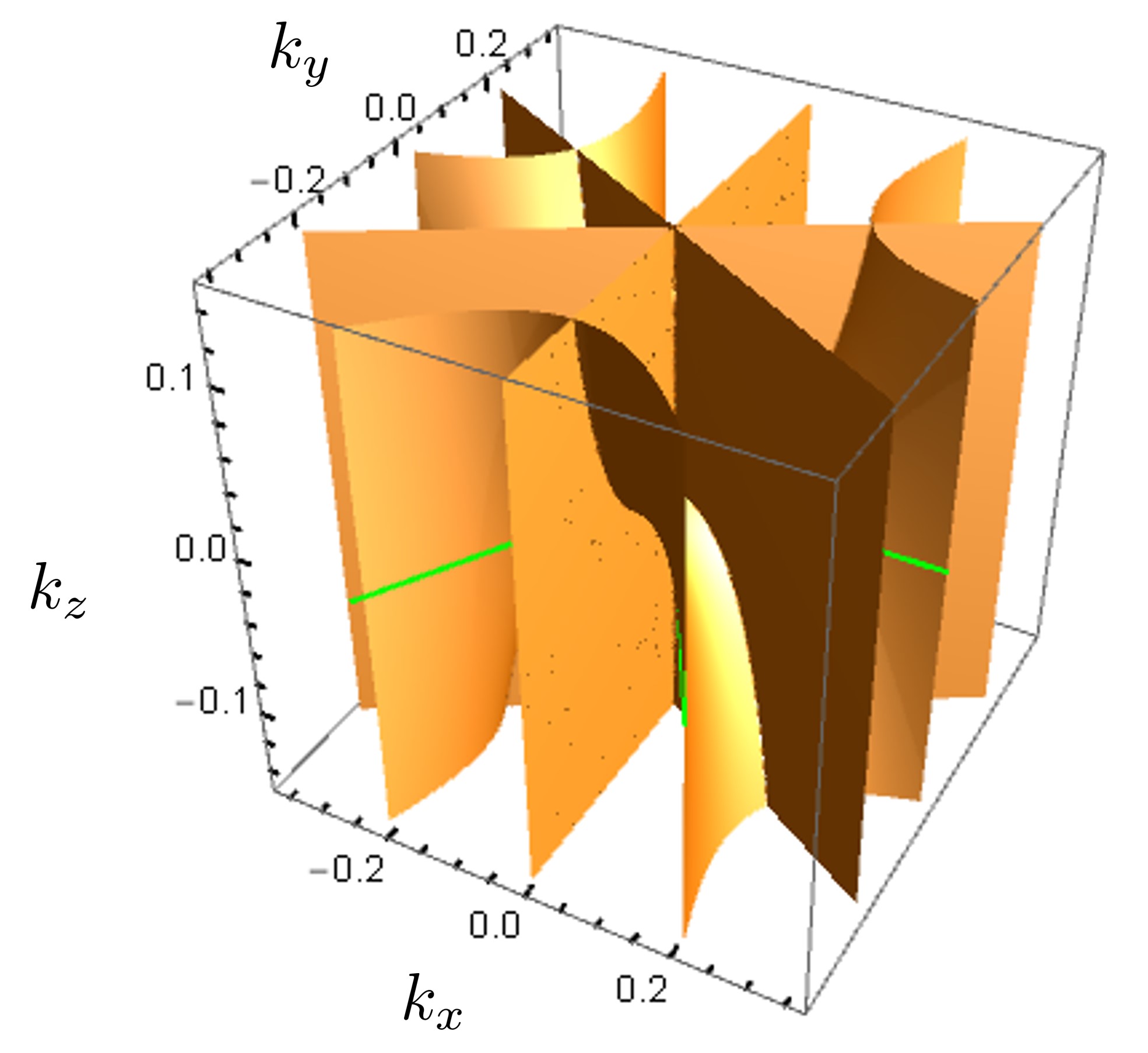}
\caption{\label{fig:nodal_surf} Nodal surfaces of $\Delta(\mathbf{k})$ in BFO near the BZ origin. The $k_x$ and $k_y$ coordinates are in units of $2 \pi / a$, while the $k_z$ coordinate is given as a fraction of the $\Gamma$-$Z$ distance. The nodal mirror planes are clearly visible, as well as the continuity-enforced nodal surface containing the symmetry-enforced two-fold axis nodal lines residing in the $k_z=0$ plane. The latter are highlighted in green.}
\end{figure}

In addition to the symmetry-enforced nodes, the heatmaps display nodal regions whose existence and location do not appear to be guaranteed by any symmetries; these are shown by dashed lines in Fig.~\ref{fig:nodal_highlight}. However, despite this lack of enforcement, these nodes are not accidental. In particular, as $k_z$ approaches zero, the BZ boundary, or the planes located one-third or two-thirds of the distance between $\Gamma$ and $Z$, it can be seen that the nodes smoothly connect to the two-fold axis nodal lines; Fig.~\ref{fig:nodal_highlight} demonstrates how this occurs in the vicinity of the $k_z=0$ plane. We note that the annular node centered on the origins of the heatmaps in the same figure is connected to the two-fold nodal axes in the other planes. As a result, the two-fold axis nodes are found to be a part of a set of interconnected nodal surfaces which appear as additional nodal lines in the two-dimensional setting of the heatmaps. A portion of one of these surfaces, which includes the BZ origin and two-fold axes in the $k_z=0$ plane, is shown in Fig.~\ref{fig:nodal_surf}, while additional and complete nodal surfaces for BFO can be observed in the SM Fig.~S1. We will refer to such nodal surfaces as ``continuity-enforced"; while their precise shapes are determined by the details of the system rather than by symmetry, their existence and connection to symmetry-enforced nodes are guaranteed by the continuity of the spin-splitting function, and they will be present in the spin-splitting function of any system with a given spin Laue group. We note that the continuity-enforced nodal surfaces for a spin-splitting function for a given system will depend in detail on the choice of bands; this is illustrated in SM Fig.~S4. 

While we do not observe accidental degeneracies in the computed heatmaps for this particular pair of bands, they are possible in general, and might be observed for other pairs of bands, or when considering multiple pairs of spin-split bands together. According to a codimension argument \cite{berry-book85,vanderbilt-book18,armitage-rmp18}, the nodal regions for the spin-splitting function have codimension one, and so will generally form two-dimensional surfaces in the three-dimensional reciprocal space. 

\subsection{Symmetry-adapted plane wave parametrization of the reciprocal space spin-splitting function}
\label{sec:shells}
To obtain a better understanding of the spin-splitting function $\Delta(\textbf{k})$ and its nodal structure, here we develop a simple parametrization of $\Delta(\textbf{k})$. Inspired by the seminal works of Refs.~\cite{chadi-prb73_1,chadi-prb73_2,baldereschi-prb73,joannopolous-jpc73,monkhorst-prb76}, we describe $\Delta(\textbf{k})$ using symmetry-adapted linear combinations of plane waves. As we will see, these functions offer a natural way of understanding the nodal structures - symmetry-enforced, continuity-enforced, and accidental - both in simple altermagnets and more complicated cases such as G-type BFO.

To begin, we further develop the discussion of the symmetry of $\Delta(\mathbf{k})$. We recall that Sec.~\ref{sec:SS-symmetries} demonstrated that the symmetry properties
of $\Delta(\textbf{k})$ are characterized by the spin Laue groups.
$\Delta(\textbf{k})$ is real and is even under inversion $[E||\mathcal{P}]$, with $\Delta(\mathbf{k})=\Delta(-\mathbf{k})$. Spin Laue group elements of the form $[E||S_j]$ and $\mathcal{T}[C_2||S_j]$ thus give $\Delta(\mathbf{k})=\Delta(S_j \mathbf{k})$ and $\Delta(\mathbf{k})=\Delta(-S_j \mathbf{k})=\Delta(S_j \mathbf{k})$, respectively, while elements of the form $[C_2||S_j]$ and $\mathcal{T}[E||S_j]$ give $\Delta(\mathbf{k})= - \Delta(S_j \mathbf{k})$ and $\Delta(\mathbf{k})= - \Delta(-S_j \mathbf{k}) = -\Delta(S_j \mathbf{k})$, respectively.  
The corresponding crystallographic Laue group is formed from the spin Laue group by considering only $\mathcal{P}$ and the real space operations $S_j$. 
It follows that $\Delta(\textbf{k})$ is a partner function of the real, parity-even, one-dimensional irreducible representation (irrep) $\alpha$ of the crystallographic Laue group, with character $-1$ for operations that are combined with either spin or time reversal, but not both, in the spin Laue group, and $+1$ otherwise.

As the band energies present in Eq.~(\ref{eq:SS-splitting}) are periodic functions of wavevector $\textbf{k}$, the spin-splitting function $\Delta(\mathbf{k})$ inherits this periodicity and can be expressed as a Fourier series: 
\begin{equation} 
\Delta(\mathbf{k}) = \sum_{\mathbf{R}} \Delta(\mathbf{R}) e^{i\mathbf{k} \cdot \mathbf{R}}, 
\label{eq:SS_Fourier} 
\end{equation} 
where $\mathbf{R}$ runs over direct lattice vectors. The constraints imposed on $\Delta(\textbf{k})$ by the spin Laue group symmetries are manifest in the Fourier coefficients $\Delta(\textbf{R})$. In particular, the $[C_2||S]$, $\mathcal{T}[C_2||E]$, and $\mathcal{T}[E||S]$ symmetries (Eqs.~(\ref{eq:SpinRot})-(\ref{eq:RotTime})) result in the Fourier coefficients satisfying 
\begin{align}
\Delta(\textbf{R})&=-\Delta(S\textbf{R}),\label{eq:SS_FR_SpinRot}\\
\Delta(\textbf{R})&=\Delta(-\textbf{R}),\label{eq:SS_FR_SpinTime}\\
\Delta(\textbf{R})&=-\Delta(-S\textbf{R}),\label{eq:SS_FR_TimeRot}
\end{align} respectively. In the context of spin Laue groups, the last two equations above are equivalently enforced by $[E||\mathcal{P}]$ and $[C_2||S\mathcal{P}]$. As $\Delta(\textbf{k})$ is real-valued, the coefficients must also satisfy $\Delta(\textbf{R})=\Delta^*(-\textbf{R})$, which combined with Eq.~(\ref{eq:SS_FR_SpinTime}) implies that they are real. 

Equation (\ref{eq:SS_Fourier}) can be re-expressed as a sum over stars, which in this case are the sets of direct lattice vectors related by operations $S$ of the crystallographic Laue group. Specifically, we construct the orthonormal symmetry-adapted combinations of plane waves $e^{i\mathbf{k} \cdot \mathbf{R}}$ transforming according to the crystallographic Laue group irrep $\alpha$: 
\begin{equation} 
W_s(\mathbf{k}) = \frac{1}{\sqrt{N_s}}\sum^{N_s}_{i=1} \chi^\alpha(S_i) e^{i\mathbf{k} \cdot \mathbf{R}_{s,i}}, 
\label{eq:SAPW}
\end{equation} where $N_s$ is the number of elements in star $s$, $\textbf{R}_{s,i}=S_i\textbf{R}_{s,1}$ is the $i$-th lattice vector in the star, $S_1=E$, and $\chi^\alpha(S_i)$ is the character of $S_i$ in the irrep $\alpha$. The lattice vectors in a star all have the same magnitude, and we order the stars by the magnitude of their vectors. We note that some stars do not contain symmetry-adapted linear combinations for the relevant $\alpha$, and thus do not contribute to $\Delta(\textbf{k})$; here we only enumerate the nontrivial stars. We then express the spin splitting function as the linear combination: \begin{equation} \Delta(\textbf{k})=\sum^{\infty}_{s=1}\Delta_sW_s(\textbf{k}). \label{eq:star_sum}\end{equation}
By construction, the symmetry-adapted plane wave functions will exhibit the same symmetries and share the same symmetry-enforced nodal regions as $\Delta(\textbf{k})$. They will also all feature continuity-enforced nodal surfaces. 

\begin{figure}[b] 
\includegraphics[width=0.48\textwidth]{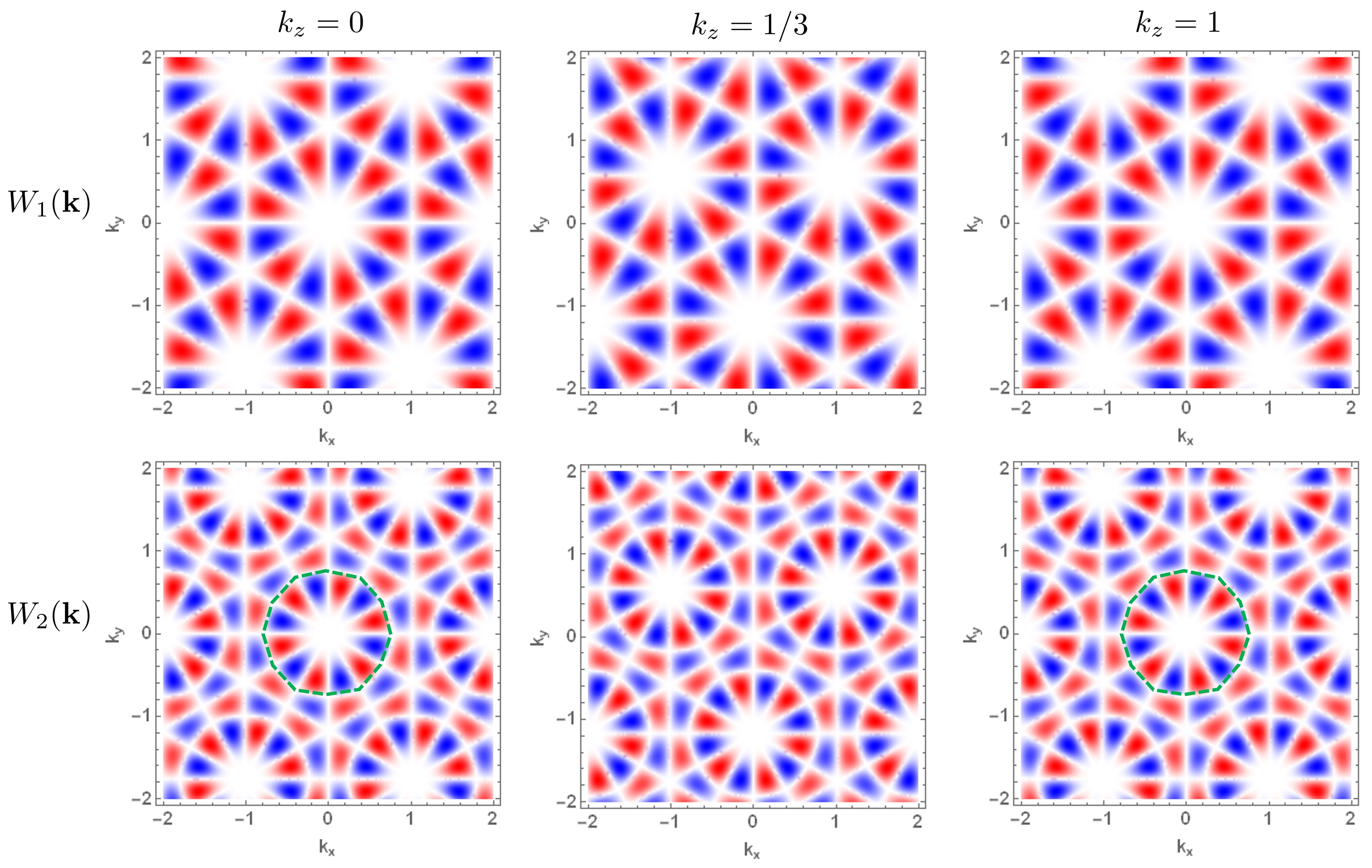}
\caption{\label{fig:BFO_W1&W2} Heatmaps at constant $k_z$ values corresponding to the symmetry-adapted linear combinations of plane waves functions $W_1(\textbf{k})$ and $W_2(\textbf{k})$ in BFO. The heatmaps in the three columns correspond to $k_z=0$, $k_z=1/3$, and $k_z=1$, where the $k_z$ values are reported as a fraction of the $\Gamma$-$Z$ distance. The top row features heatmaps generated for $W_1(\textbf{k})$, while the bottom row is for $W_2(\textbf{k})$. The $k_x$ and $k_y$ coordinates are in units of $2\pi/a$, with $a=5.668$ \AA\ being the lattice constant of the rhombohedral cell.}
\end{figure}

\begin{figure}[b] 
\includegraphics[width=0.48\textwidth]{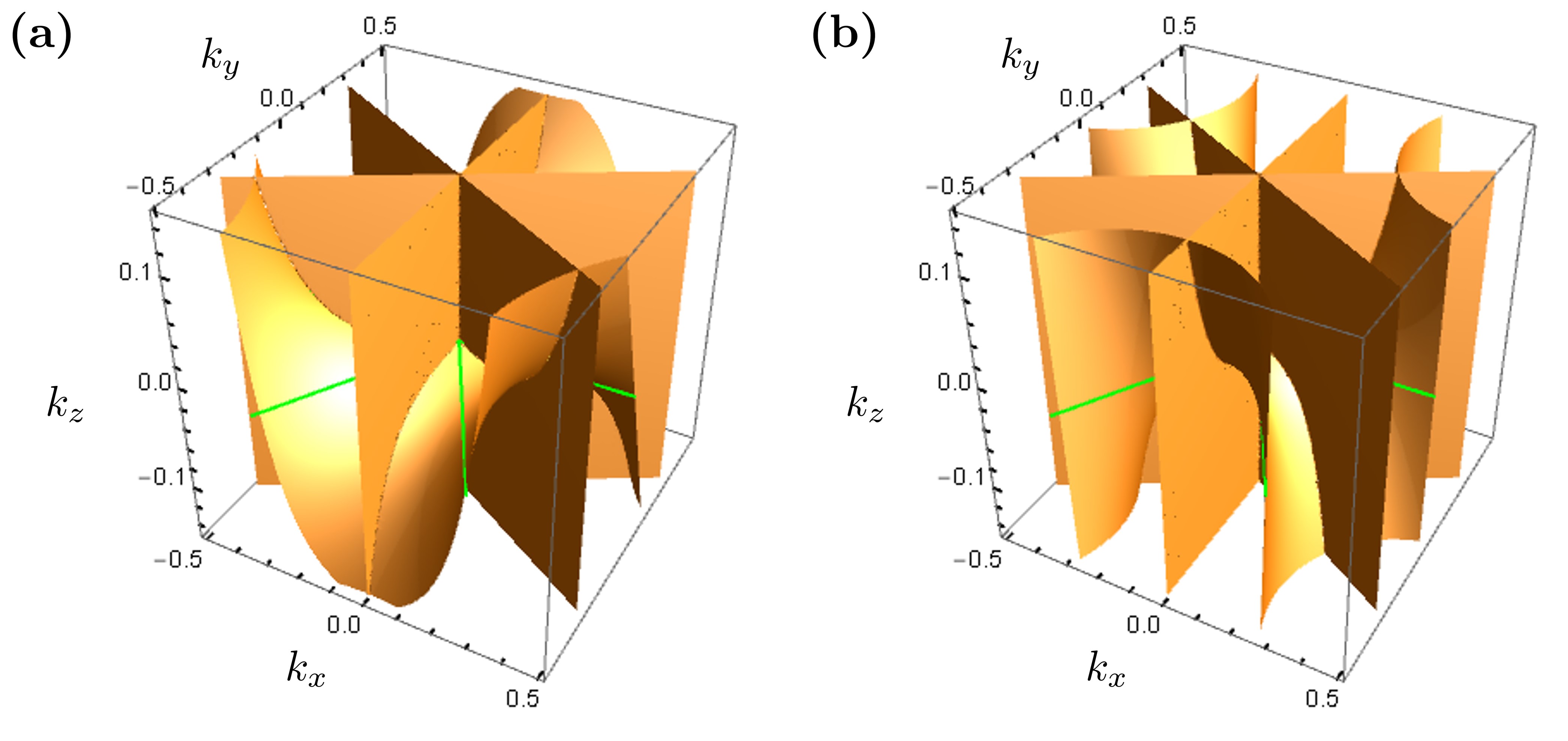}
\caption{\label{fig:BFO_W1&W2_nodes} The nodal surfaces near the BZ origin for (a) $W_1(\textbf{k})$ and (b) $W_2(\textbf{k})$. The $k_x$ and $k_y$ coordinates are in units of $2 \pi / a$, while the $k_z$ coordinate is given as a fraction of the $\Gamma$-$Z$ distance.  The three symmetry-enforced mirror nodal planes and two-fold axis nodal lines are present, with the latter colored in green. The nodal axis lines are smoothly connected to continuity-enforced nodal surfaces, which differ in appearance between the two functions.}
\end{figure}

In the specific case of BFO, since the mirror operations are combined with spin reversal and the three-fold rotation is not, $\Delta(\textbf{k})$ transforms under the $\Gamma^+_2$ irrep of the $\bar{3}m$ crystallographic Laue group. The first two symmetry-adapted plane wave functions for BFO are then
\begin{widetext}
\begin{align} 
W_1(\mathbf{k}) = (1/\sqrt{3}) \{&\cos\left[\mathbf{k} \cdot (2 \mathbf{a}_1 - \mathbf{a}_2)\right] + 
\cos\left[\mathbf{k} \cdot (2 \mathbf{a}_2 - \mathbf{a}_3)\right] + \cos\left[\mathbf{k} \cdot (2 \mathbf{a}_3 - \mathbf{a}_1)\right] \nonumber \\ - &\cos\left[\mathbf{k} \cdot (2 \mathbf{a}_2 - \mathbf{a}_1)\right] - 
\cos\left[\mathbf{k} \cdot (2 \mathbf{a}_3 - \mathbf{a}_2)\right] -
\cos\left[\mathbf{k} \cdot (2 \mathbf{a}_1 - \mathbf{a}_3)\right]
\},
\end{align}
\begin{align}
W_2(\textbf{k}) = (1/\sqrt{3}) \{&\cos\left[\mathbf{k} \cdot (2 \mathbf{a}_1 - 2 \mathbf{a}_2 + \mathbf{a}_3)\right] + \cos\left[\mathbf{k} \cdot (2 \mathbf{a}_2 - 2 \mathbf{a}_3 + \mathbf{a}_1)\right] + \cos\left[\mathbf{k} \cdot (2 \mathbf{a}_3 - 2 \mathbf{a}_1 + \mathbf{a}_2)\right] \nonumber \\ 
- &\cos\left[\mathbf{k} \cdot (2 \mathbf{a}_2 - 2 \mathbf{a}_1 + \mathbf{a}_3)\right] - \cos\left[\mathbf{k} \cdot (2 \mathbf{a}_3 - 2 \mathbf{a}_2 + \mathbf{a}_1)\right] - \cos\left[\mathbf{k} \cdot (2 \mathbf{a}_1 - 2 \mathbf{a}_3 + \mathbf{a}_2)\right]  \},
\end{align} 
\end{widetext}
where $\textbf{a}_1$, $\textbf{a}_2$, and $\textbf{a}_3$ are the primitive lattice vectors. We note that the ordering of the $W_s(\textbf{k})$ functions depends on the value of the angle between the primitive lattice vectors, which for BFO is $59.06^\circ$. Figure~\ref{fig:BFO_W1&W2} shows heatmaps of $W_1(\textbf{k})$ and $W_2(\textbf{k})$, plotted on reciprocal space slices at $k_z = 0$, $1/3$, and $1$ (reported as the fraction of the $\Gamma$-$Z$ distance). The mirror nodal planes and two-fold axis nodal lines are clearly visible in all slices. These nodal lines are contained in the continuity-enforced nodal surfaces, which are shown in more detail in panels (a) and (b) of Fig.~\ref{fig:BFO_W1&W2_nodes} for $W_1(\mathbf{k})$ and $W_2(\mathbf{k})$ respectively, and in the heatmaps in SM Figs.~S2 and S3. Although the surfaces differ in shape between the two functions, both respect the symmetries of BFO. Their specific geometries are directly determined by the structure of the primitive unit cell. Additional nodal features may also be observed in the heatmaps — for instance, the green-highlighted regions in the $k_z = 0$ and $k_z = 1$ slices of $W_2(\mathbf{k})$ — but these are not accidental; rather, they are part of the continuity-enforced surface. This can be seen in detail in the SM Fig.~S3. Similar continuity-enforced nodal surfaces appear in all higher-order $W_s(\textbf{k})$. 

For well-behaved $\Delta(\textbf{k})$, the coefficients $\Delta_s$ vanish in the limit $s\to\infty$, allowing the sum in Eq.~(\ref{eq:star_sum}) to be truncated at sufficiently large $s$ while still accurately approximating the spin-splitting function. In BFO, we constructed such an approximation by performing a least-squares fit of the $\Delta_s$ coefficients to the spin-splitting data obtained from DFT. Our fitting was performed for the first nine nontrivial stars, and the second star was found to be the leading contribution to the spin splitting. For further details on the values of the fitting coefficients, as well as a representative lattice vector for each nontrivial star, we refer the reader to the SM. The plot of Fig.~\ref{fig:nodal_surf} is obtained from the resulting fit to the spin-splitting function.

Monkhorst and Pack \cite{monkhorst-prb76} demonstrated that symmetry-adapted linear combinations of plane waves transforming under the identical irrep of the crystallographic point group are orthonormal on special uniform sets of $\textbf{k}$-points. In particular, the set of such symmetry-adapted linear combinations of plane waves will be orthonormal on a Monkhorst-Pack mesh of $N^3_\textbf{k}$ uniformly spaced $\textbf{k}$-points, provided that the internal coordinates of the lattice vectors in the corresponding stars are simultaneously bounded from below and above by $-N_\textbf{k}/2$ and $N_\textbf{k}/2$, respectively. Such $\textbf{k}$-point meshes and plane waves provide a computationally efficient scheme for computing the averages of functions defined on the BZ. We note here that this argument can be readily extended to symmetry-adapted combinations of plane waves that transform, like $\Delta(\textbf{k})$, under one-dimensional inversion-even, non-identical irreps of the crystallographic Laue group. In particular, this implies that the set of $W_s(\textbf{k})$ in Eq.~(\ref{eq:SAPW}) is orthonormal on the Monkhorst-Pack mesh. So, under the assumption that $\Delta_s$ above a given $s$ is negligible, the $\Delta_s$ can be obtained by sampling the DFT results on a Monkhorst-Pack mesh. We report and compare the $\Delta_s$ obtained from a $7\times7\times7$ Monkhorst-Pack mesh to those from the least-squares fitting in the SM.

\begin{table*}[t!]
\caption{Characterization of altermagnets according to the irreps of the crystallographic Laue group that describe the spin-splitting function.}
\centering
\begin{tabular}{|c|c|c|c|}
\hline
Laue Group & One-dimensional, non-identical, & Spin Laue group & Altermagnetic \\ 
$\mathcal{G}_\mathcal{L}$ & inversion-even irrep $\mathcal{D}^{\mathcal{G}_\mathcal{L}}$ & $\mathcal{G}^{(\text{s})}_\mathcal{L}$ & pattern \\
\hline
$\bar{1}$ & None & ${}^1\bar{1}$ & No altermagnetism \\ 
\hline
$2/m$ & $\Gamma_2^{+}$ & ${}^22/{}^2m$ & Bulk $d$-wave  \\ 
\hline
\multirow{3}{*}{$mmm$} & $\Gamma_2^{+}$ & ${}^1m^2m^2m$ & Bulk $d$-wave \\ 
\cline{2-4} 
& $\Gamma_3^{+}$ & ${}^2m^1m^2m$ & Bulk $d$-wave \\
\cline{2-4} 
& $\Gamma_4^{+}$ & ${}^2m^2m^1m$ & Planar $d$-wave \\
\hline
$4/m$ & $\Gamma_2^{+}$ & ${}^24/{}^1m$ & Planar $d$-wave \\ 
\hline
\multirow{3}{*}{$4/mmm$} & $\Gamma_2^{+}$ & ${}^14/{}^1m^2m^2m$ & Planar $g$-wave \\ 
\cline{2-4}
& $\Gamma_3^{+}$ & ${}^24/{}^1m^1m^2m$ & Planar $d$-wave \\
\cline{2-4}
& $\Gamma_4^{+}$ & ${}^24/{}^1m^2m^1m$ & Planar $d$-wave \\
\hline
$\bar{3}$ & None & ${}^1\bar{3}$ & No altermagnetism \\ 
\hline
$\bar{3}m$ & $\Gamma_2^{+}$ & ${}^1\bar{3}^2m$ & Bulk $g$-wave \\ 
\hline 
$6/m$ & $\Gamma_2^{+}$ & ${}^26/{}^2m$ & Bulk $g$-wave \\ 
\hline
\multirow{3}{*}{$6/mmm$} & $\Gamma_2^{+}$ & ${}^16/{}^1m^2m^2m$ & Planar $i$-wave \\ 
\cline{2-4}
& $\Gamma_3^{+}$ & ${}^26/{}^2m^2m^1m$ & Bulk $g$-wave \\ 
\cline{2-4}
& $\Gamma_4^{+}$ & ${}^26/{}^2m^1m^2m$ & Bulk $g$-wave \\ 
\hline
$m\bar{3}$ & None & ${}^1m^1\bar{3}$ & No altermagnetism \\
\hline
$m\bar{3}m$ & $\Gamma_2^{+}$ & ${}^1m^1\bar{3}^2m$ & Bulk $i$-wave \\
\hline
\end{tabular}
\label{tab:classification}
\end{table*}

\section{\texorpdfstring{Classification of spin-splitting pattern into ($d$, $g$, $i$)-wave scheme}{g-wave}}\label{SS-classification}

Next, we address the classification of altermagnetic ordering into the bulk/planar ($d$,$g$,$i$)-wave scheme proposed in Ref.~\cite{smejkal-prx22}. There, they present a table (Fig.~2 of Ref.~\cite{smejkal-prx22}) in which the bulk or planar $d$, $g$ or $i$-wave type follows from the spin Laue group, and give a related prescription that identifies the integer $l$ distinguishing $d$-wave ($l = 2$), $g$-wave ($l = 4$), and $i$-wave ($l = 6$) altermagnets as the number of spin-degenerate nodal surfaces crossing the $\Gamma$-point.  According to that table, given that the spin point group of BFO is ${}^13^2m$, which belongs to the spin Laue group ${}^1\bar{3}^2m$, BFO is classified as a bulk $g$-wave altermagnet. However, from the examination of the spin splitting in the $k_z = 0$ plane (Fig.~\ref{fig:nodal_highlight}(a)), it may appear that there are six nodal surfaces crossing $\Gamma$, leading to the conclusion that $l = 6$ and BFO is an $i$-wave altermagnet \cite{smejkal-arxiv24}. This paradox can be resolved by noting that, as explained in the previous section and shown in Fig.~\ref{fig:nodal_surf}, the three two-fold nodal lines passing through $\Gamma$ are all part of a single interconnected nodal surface. Thus, there is a total of four nodal surfaces crossing at $\Gamma$, making BFO a $g$-wave altermagnet, consistent with the table in Ref.~\cite{smejkal-prx22}.  

This confusion led us to revisit the question of the classification of altermagnetic systems as $d$, $g$ or $i$-wave. Here, we show how to do the classification through a purely group-theoretical analysis of the spin-splitting function $\Delta(\mathbf{k})$ defined in the previous section.   
 
As discussed above in Sec.~\ref{sec:shells}, $\Delta(\mathbf{k})$ is invariant under the operations of the spin Laue group $\mathcal{G}_\mathcal{L}^{(\text{s})}$.
In addition, $\Delta(\mathbf{k})$ can be classified using the irreps of the crystallographic Laue group $\mathcal{G}_\mathcal{L}$. In particular, in collinear altermagnets, $\Delta(\mathbf{k})$ is odd under symmetry operations of $\mathcal{G}_\mathcal{L}$ that are coupled to a spin-space rotation in $\mathcal{G}_\mathcal{L}^{(\text{s})}$, while it is even under symmetries that are not coupled to a spin-space rotation in $\mathcal{G}_\mathcal{L}^{(\text{s})}$. Thus, $\Delta(\mathbf{k})$ must transform according to a one-dimensional, inversion-even, non-identical irrep of $\mathcal{G}_\mathcal{L}$ \footnote{$\Delta(\mathbf{k})$ transforms as the total-symmetric irrep of $\mathcal{G}_\mathcal{L}$ only in ferromagnets.}. The relevant irreps for each of the 11 crystallographic Laue groups are listed in Table \ref{tab:classification} \cite{Aroyo-zkri2006,Aroyo-ACA2006,Aroyo-bcc2011}. 

\begin{table}[b]
\centering
\caption{Character table of the $l$-dimensional, inversion-even (i.e., with even $l$), irreps of $O(3)$. $ C_{2 \pi / \alpha}$ identifies rotations of angle $\alpha$ about any axis.}
\begin{tabular}{|c|c|c|c|c|}
\hline
\multirow{2}{*}{Representation} & \multicolumn{4}{c|}{Symmetry operation} \\
\cline{2-5}
& $E$ & $C_{2 \pi / \alpha}$ & $\mathcal{P}$ & $\mathcal{P} C_{2 \pi / \alpha}$ \\
\hline
$\mathcal{D}^{(l)}$ & $2l + 1$ & $\sum_{m = -l}^l e^{i m \alpha}$ & $2l + 1$ & $\sum_{m = -l}^l e^{i m \alpha}$ \\
\hline
\end{tabular}
\label{tab:characters}
\end{table}

For each irrep, to determine the altermagnetic type, we consider the irreps $\mathcal{D}^{(l)}$ of the full orthogonal group $O(3)$ (Table \ref{tab:characters}) and their decomposition into irreps of $\mathcal{G}_\mathcal{L}$. To be more precise, $\mathcal{D}^{(l)}$ is a \textit{reducible} representation of $\mathcal{G}_\mathcal{L}$ and can be written as a direct sum of irreps of $\mathcal{G}_\mathcal{L}$ 
\begin{equation}
    \mathcal{D}^{(l)} = \bigoplus_{i = 1}^{N_c} n_i \mathcal{D}_i^{\mathcal{G}_\mathcal{L}},
\end{equation}
where $N_c$ is the number of irreps of $\mathcal{G}_\mathcal{L}$. Each irrep $\mathcal{D}_i^{\mathcal{G}_\mathcal{L}}$ appears $n_i$ times in the decomposition of $\mathcal{D}^{(l)}$. The integer $n_i$ is readily computed from the characters of the representations as follows: 
\begin{equation}
    n_i = \frac{1}{|\mathcal{G}_\mathcal{L}|} \sum_S \chi^{(i)}_{\mathcal{G}_\mathcal{L}}(S)^* \chi_l(S). 
    \eqlab{ni}
\end{equation}
Above, $S$ runs over the elements of $\mathcal{G}_\mathcal{L}$. $|\mathcal{G}_\mathcal{L}|$ is the number of elements in the same group, and $\chi^{(i)}_{\mathcal{G}_\mathcal{L}}$ and $\chi_l$ are the characters of the $\mathcal{D}_i^{\mathcal{G}_\mathcal{L}}$ and $\mathcal{D}^{(l)}$ representations, respectively. The characters of $\mathcal{D}^{(l)}$ are given for the different types of symmetry operations (identity, inversion, rotations, roto-inversions) in Table \ref{tab:characters} (see also Chapter 1 of Ref.~\cite{bassani-75}). Once $n_i$ has been computed from Eq.~\eqref{eq:ni}, we define the altermagnetic integer as the lowest value of $l$ such that $\mathcal{D}^{(l)}$ contains the irrep of $\Delta(\mathbf{k})$ at least once. 

For example, $\Delta(\mathbf{k})$ for BFO transforms according to the $\Gamma_2^{+}$ irrep of the crystallographic Laue group $\bar{3}m$. Applying \eq{ni} to the representations $\mathcal{D}^{(l)}$ with $l = 2, 4, 6$ gives the following decompositions of $\mathcal{D}^{(l)}$ into irreps of $\bar{3}m$: 
\begin{align}
\mathcal{D}^{(2)} & = \Gamma_1^{+} \oplus 2 \Gamma_3^{+}, \\
\mathcal{D}^{(4)} & = 2 \Gamma_1^{+} \oplus \Gamma_2^{+} \oplus 3 \Gamma_3^{+}, \\
\mathcal{D}^{(6)} & = 3 \Gamma_1^{+} \oplus 2 \Gamma_2^{+} \oplus 4 \Gamma_3^{+}. \end{align}
The lowest $l$ for which $\Gamma_2^{+}$ appears in the decomposition of $\mathcal{D}^{(l)}$ is $l = 4$, and thus the altermagnetic type is $g$-wave. We further classify it as bulk since the character of the $\Gamma_2^{+}$ irrep is $-1$ for the two-fold axes in the $k_z$=0 plane.

The complete results for all the crystallographic Laue group irreps are presented in Table \ref{tab:classification}. For each one-dimensional, inversion-even, nontrivial irrep of $\mathcal{G}_\mathcal{L}$, we report the corresponding spin Laue group and the classification of the altermagnetic pattern under the ($d$,$g$,$i$)-wave scheme following from the characteristic integer $l$. Among the 11 crystallographic Laue groups, three ($\bar{1}$, $\bar{3}$, and $m\bar{3}$) do not support altermagnetism, because they do not have a one-dimensional, non-identical, inversion-even irrep. Finally, we note that our table completes the table in Fig.~2 of Ref.~\cite{smejkal-prx22} by the inclusion of four additional Laue group irreps, corresponding to spin Laue groups ${}^2m^1m^2m$, ${}^1m^2m^2m$, ${}^24/{}^1m^1m^2m$, ${}^26/{}^2m^1m^2m$. The first two can be obtained from ${}^2m^2m^1m$ by a change of setting of the orthorhombic axes, though it should be noted that classification as bulk or planar depends on the setting. For the latter two, the nodal surfaces are rotated relative to the reciprocal lattice vectors by 45 and 30 degrees, respectively, from those for ${}^24/{}^1m^2m^2m$ and ${}^26/{}^2m^2m^1m$. 

\section{Control of the spin splitting via structural distortions}
\label{sec:struct-dist}

Recently, the control of altermagnetic spin splitting through applied electric field or stress has been investigated, focusing on ferroelectrics \cite{gu-prl25,smejkal-arxiv24,dong-arxiv25}. For an altermagnetic ferroelectric, the paraelectric parent phase contains $\mathcal{PT}$ symmetry, $U\mathbf{t}$ symmetry or both, establishing three distinct classes. A zone-center polar mode always breaks $\mathcal{PT}$ but cannot break $U\mathbf{t}$. So, if $U\mathbf{t}$ is present in the parent phase, there must also be a zone-boundary mode distortion, typically characterized by tilts of the ligand polyhedra surrounding the magnetic ions. 

For G-type BFO, we take the ideal cubic perovskite structure (space group P$m\bar{3}m$) as a paraelectric parent phase. 
The checkerboard pattern of G-type AFM ordering means that in the high-symmetry cubic phase, any two neighboring cells along the [111] direction belong to the two opposite magnetic sublattices. As a consequence, the two sublattices are connected by $U \mathbf{t}$ symmetry, with $\mathbf{t} = (1/2, 1/2, 1/2)$. G-type P$m\bar{3}m$ BFO also has $\mathcal{PT}$ symmetry, with the inversion center on the Bi site.

\begin{figure}[t] 
\includegraphics[width=0.5\textwidth]{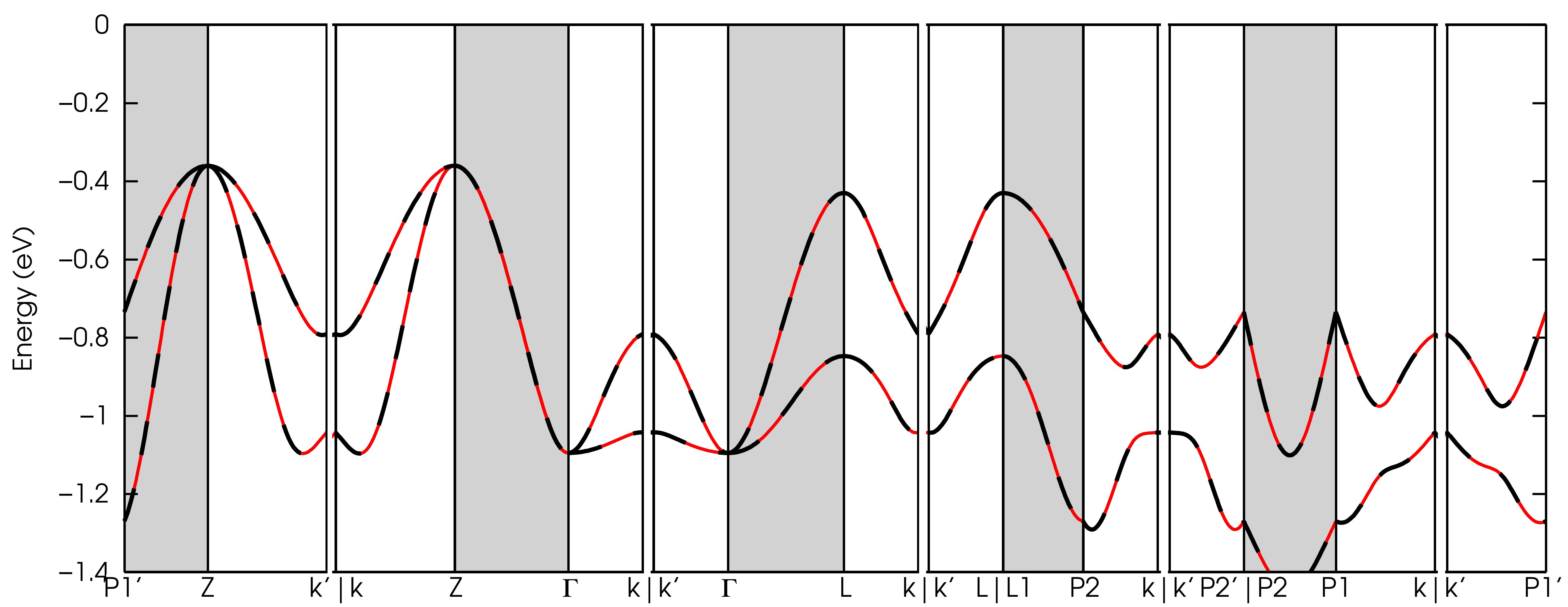}
\caption{\label{fig:bands_R3m} The top four valence bands in the band structure of BFO in the R$3m$ structure obtained by condensing the $\Gamma_4^-$ mode of the parent P$m\bar{3}m$ phase.}
\end{figure}

\begin{figure}[t] 
\includegraphics[width=0.5\textwidth]{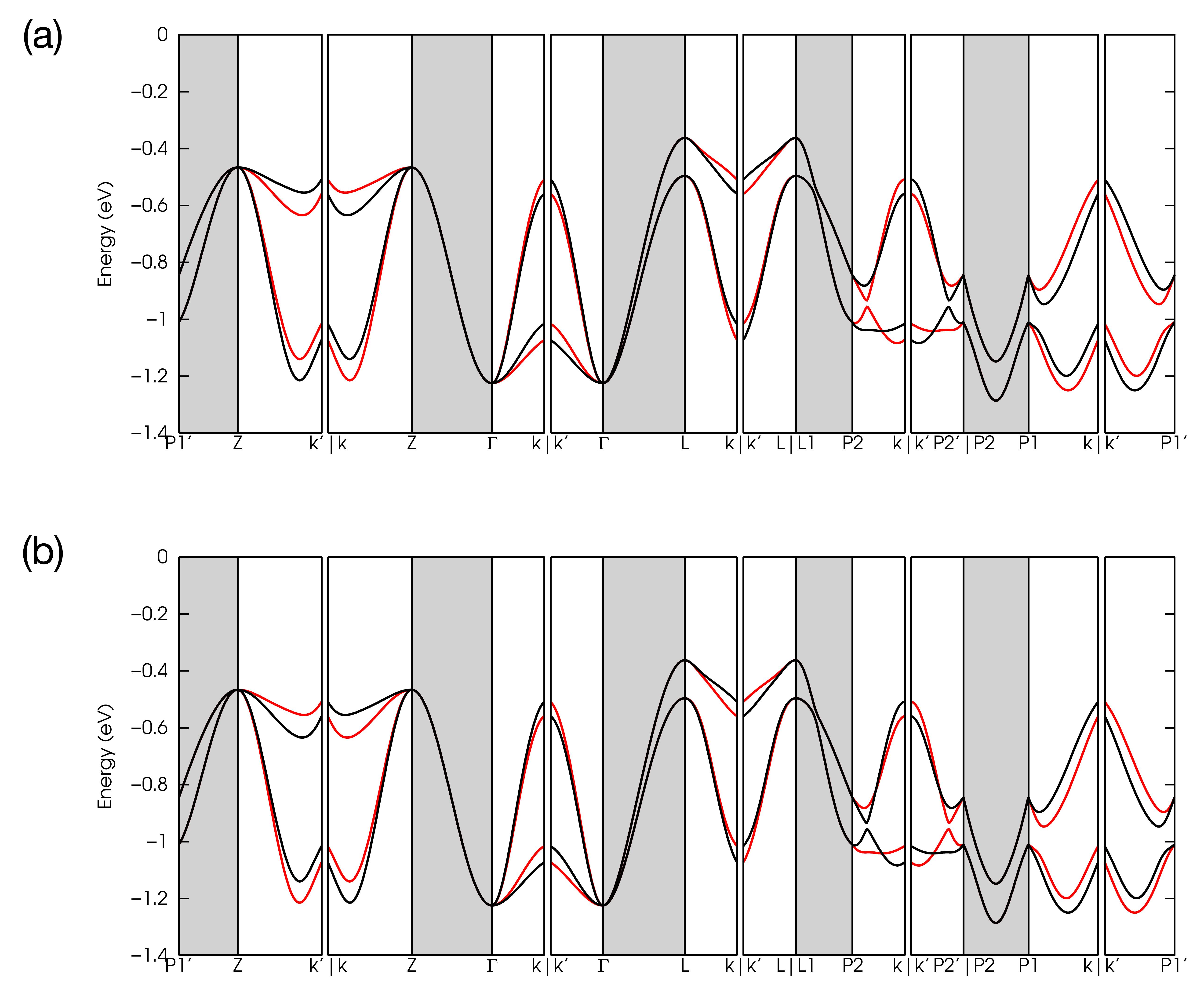}
\caption{\label{fig:bands_R-3c} (a) The top four valence bands in the band structure of BFO in the R$\bar{3}c$ structure, obtained by condensing the $R_4^+$ mode of the parent P$m\bar{3}m$ phase. (b) Same as (a) except that the sign of the $R_4^+$ mode is reversed.}
\end{figure}

As discussed in Sec.~\ref{sec:symm}, the R$3c$ structure is obtained from P$m\bar{3}m$ by the superposition of two unstable modes: (1) a zone-boundary $R_4^+$ mode, characterized by oxygen octahedron rotations around the [111] direction, and (2) a zone-center polar mode $\Gamma_4^-$, leading to polarization along the [111] direction. 
The $\Gamma_4^-$ polar mode alone leads to a non-centrosymmetric phase R$3m$. Here, the two magnetic sublattices are still connected by the $(1/2, 1/2, 1/2)$ fractional translation, and thus the R$3m$ phase is not altermagnetic. This is confirmed by Fig.~\ref{fig:bands_R3m}, where we show the top four valence bands (two spin-up, two spin-down). 
On the other hand, the $R_4^+$ mode alone leads to a centrosymmetric R$\bar{3}c$ structure, which unlike R$3m$ breaks the fractional translation connecting the two sublattices, because the rotations of the oxygen octahedra are opposite for the two sublattices. As a consequence, $U\mathbf{t}$ symmetry is broken.
In addition, while the  $R_4^+$ mode preserves the inversion center on Fe, it breaks the inversion center on Bi, and thus breaks the $\mathcal{PT}$ symmetry. Thus the R$\bar{3}c$ phase is altermagnetic. In Fig.~\ref{fig:bands_R-3c} we show the top four valence bands for two configurations with opposite oxygen octahedra rotations. As expected, the spin splitting is opposite in the two configurations.

The possibility of switching altermagnetic spin splitting by electric-field switching in BFO therefore hinges on whether the oxygen octahedron rotations switch when the polarization is reversed. The process of electric-field switching in BFO has been much discussed, and it has been proposed that BFO does switch in this way~\cite{heron-nature14}.

BFO is thus an example of a ferroelectric that has both $U\mathbf{t}$ and $\mathcal{PT}$ symmetries in the parent paraelectric phase. To be able to switch altermagnetism directly by switching the polar mode, we should consider magnetic ferroelectrics where the parent paraelectric nonaltermagnetic phase has only $\mathcal{PT}$ symmetry and not $U\mathbf{t}$ symmetry. 

\section{Conclusions}
\label{sec:conclusions}

In summary, based on symmetry analysis and on \textit{ab initio} calculations, we have shown that G-type AFM BFO is a $g$-wave altermagnet. Through our investigation of the particular case of BFO, we have developed several tools and concepts for the broader study of altermagnetism. To make the spin splitting evident, we introduced a bandstructure plotting scheme with a generalized path in reciprocal space that includes both high-symmetry and general lines, and highlights in a clear way the spin splitting and its altermagnetic nature. This bandstructure plotting scheme should be illuminating for any altermagnetic system, and has advantages for the investigation of electronic states and resulting physical properties in any crystal.

We confirmed the $g$-wave character of the spin splitting in BFO through a group-theoretical approach. We investigated the detailed behavior of the spin-splitting function, defined for an isolated pair of valence bands of BFO, by showing heatmaps that sample the entire BZ. We analyzed the nodal surfaces of the spin-splitting function and showed that the symmetry-related surfaces are either symmetry-enforced mirror planes or are smooth surfaces that contain two-fold rotation axes of the crystallographic Laue group. We introduced a generalized scheme to parametrize the spin-splitting function, using sets of symmetry-adapted linear combinations of plane waves. Our approach applies to all lattice-periodic functions in reciprocal space that transform according to the same irrep of the crystallographic Laue group as the spin-splitting function, and the approach can be further generalized to functions that transform according to a different irrep of the crystallographic Laue group.  

Finally, we showed that the spin splitting of BFO is controlled by the rotation of oxygen octahedra. Since the octahedral rotations alone do not generate electric polarization, it might seem that altermagnetism in BFO is not controllable by an external electric field. However, as shown in Ref.~\cite{heron-nature14}, the octahedral rotations in BFO are in fact locked to the polar mode, allowing switching of the weak ferromagnetic moment by an electric field. By the same token, we expect the altermagnetic spin splitting to be electric-field switchable. We also note that direct switching by an electric field should be possible in altermagnetic ferroelectrics for which the parent nonaltermagnetic paraelectric phase has only $\mathcal{PT}$ symmetry and not $U\mathbf{t}$ symmetry.

\section*{Acknowledgments}

The authors thank David Vanderbilt, Nicola A.~Spaldin, and Sang-Wook Cheong for useful discussions. A.U. acknowledges support from the Abrahams Postdoctoral Fellowship of the Center for Materials Theory at Rutgers University. Y.T. acknowledges support from the Wisconsin Materials Research Science and Engineering Center (NSF DMR-2309000). D.S. and K.M.R. acknowledge support from Office of Naval Research N00014-21-1-2107. S.E.R.-L. acknowledges ANID Fondecyt regular grant number 1220986. Part of the work by K.M.R. was done on a sabbatical visit to the group of Annabella Selloni and Roberto Car at Princeton University, and at the Aspen Center for Physics, which is supported by the National Science Foundation grant PHY-2210452. S.Y.P was supported by the National Research Foundation of Korea (NRF) grant funded by the Korea government (MSIT) (RS-2024-00358551). Computational facilities were provided by the Beowulf cluster at the Department of Physics and Astronomy of Rutgers University.

\bibliography{cite}

\end{document}



\title{\texorpdfstring{Supplementary Material for ``G-type Antiferromagnetic 
BiFeO$_3$ is a Multiferroic $g$-wave Altermagnet''\\
}{BFO-title}}

\author{Andrea Urru}
\thanks{These authors contributed equally to the work.}
\affiliation{Department of Physics $\&$ Astronomy, Center for Materials Theory, Rutgers University, 
Piscataway, New Jersey 08854, United States}%

\author{Daniel Seleznev}
\thanks{These authors contributed equally to the work.}
\affiliation{Department of Physics $\&$ Astronomy, Center for Materials Theory, Rutgers University, 
Piscataway, New Jersey 08854, United States}%

\author{Yujia Teng}
\affiliation{Department of Physics $\&$ Astronomy, Center for Materials Theory, Rutgers University, 
Piscataway, New Jersey 08854, United States}%

\author{Se Young Park}
\affiliation{Department of Physics and Origin of Matter and Evolution of Galaxies (OMEG) Institute, Soongsil University, Seoul, Korea}

\author{Sebastian E. Reyes-Lillo}
\affiliation{Departamento de F\'isica y Astronom\'ia, Universidad Andres Bello, Santiago 837-0136, Chile}

\author{Karin M. Rabe}
\affiliation{Department of Physics $\&$ Astronomy, Center for Materials Theory, Rutgers University, 
Piscataway, New Jersey 08854, United States}


\maketitle

\section{Computational details}
\label{app:comp-details}

\textit{Ab initio} DFT calculations are performed within the generalized gradient approximation (GGA), as implemented in \texttt{Quantum ESPRESSO}~\cite{giannozzi-jpcm09,giannozzi-jpcm17}, with the Perdew-Burke-Ernzerhof (PBE) scheme~\cite{perdew-prl96} to treat the exchange-correlation energy. Ions are described using scalar-relativistic and full-relativistic optimized norm-conserving Vanderbilt pseudopotentials (ONCV PPs)~\cite{hamann-prb13}, with valence configurations 5$d^{10}$6$s^2$6$p^3$ for Bi, 3$s^2$3$p^6$4$s^2$3$d^6$ for Fe, and 2$s^2$2$p^4$ for O. Magnetic order is treated at the collinear (non-collinear) level, i.e., with scalar (spinor) wave functions, for calculations without (with) spin-orbit coupling included. The pseudo-wave functions are expanded in a plane-wave basis set with kinetic energy cut-off of 90 Ry. Brillouin Zone integrations are performed on a 6$\times$6$\times$6 $\Gamma$-centered Monkhorst-Pack mesh~\cite{monkhorst-prb76}. Self-consistent ground-state calculations are performed with a convergence threshold of $10^{-10}$ Ry. Structural relaxations are performed with a convergence threshold of $10^{-3}$ Ry/a.u.\ on the forces. The relaxed lattice constants and atomic positions parameters are reported in Table \ref{tab:structure}, both in the non-conventional (rhombohedral) and conventional (hexagonal) settings. In the same table, reference experimental values of the structural parameters~\cite{kubel-acb90} are also reported in parentheses. Band structures to obtain the spin-splitting maps (Figs.\ 4 and 5 in the main text and Figs.~\ref{figs1} and \ref{figs4} below) are computed with non-self-consistent calculations in a uniform 31$\times$31$\times$31 $\Gamma$-centered mesh. The fractional coordinates (in the rhombohedral setting) of the high-symmetry points appearing in the generalized path in reciprocal space used for Figs.\ 3, 9, and 10 in the main text are given in Table \ref{tab:kpoints}.

\begin{table}[h]
\centering
\caption{BFO \textit{Ab initio} equilibrium lattice constants and atomic positions, given in both the rhombohedral and hexagonal settings. Experimental reference values \cite{kubel-acb90} are reported in parentheses.}
\begin{tabular}{|c|c c c||c|c c c|}
\hline
\multicolumn{4}{|c||}{Rhombohedral setting} & \multicolumn{4}{|c|}{Hexagonal setting} \\
\hline 
$a$ (\AA) & \multicolumn{3}{|c||}{5.67 (5.63)} & $a$ (\AA) & \multicolumn{3}{|c|}{5.59 (5.57)} \\
$\alpha$ (deg) & \multicolumn{3}{|c||}{59.06 (59.35)} & $c$ (\AA) & \multicolumn{3}{|c|}{13.98 (13.86)} \\
\hline 
\hline
\multirow{2}{*}{Bi (2a)} & $0.000$ & $0.000$ & $0.000$ & \multirow{2}{*}{Bi (6a)} & \multirow{2}{*}{0} & \multirow{2}{*}{0} & $0.000$ \\
& (0.000) & (0.000) & (0.000) & & & & (0.000) \\
\hline
\multirow{2}{*}{Fe (2a)} & 0.226 & 0.226 & 0.226 & \multirow{2}{*}{Fe (6a)} & \multirow{2}{*}{0} & \multirow{2}{*}{0} & 0.226 \\ 
& (0.221) & (0.221) & (0.221) & & & &  (0.221) \\
\hline
\multirow{2}{*}{O (6b)} & 0.943 & 0.382 & 0.545 & \multirow{2}{*}{O (18b)} & 0.429 & 0.018 & 0.957 \\
& (0.933) & (0.395) & (0.528) & & (0.443) & (0.019) & (0.952) \\
\hline
\end{tabular}
\label{tab:structure}
\end{table}


\begin{table}
\centering
\caption{High-symmetry points of BFO, specified in reduced coordinates, $\mathbf{k} = h \mathbf{b}_1 + k \mathbf{b}_2 + l \mathbf{b}_3$, in the rhombohedral setting.}
\begin{tabular}{|c|c|c|c|c|c|c|c|}
\hline
High-symmetry & \multirow{2}{*}{$h$} & \multirow{2}{*}{$k$} & \multirow{2}{*}{$l$} & High-symmetry & \multirow{2}{*}{$h$} & \multirow{2}{*}{$k$} & \multirow{2}{*}{$l$} \\
point & & & & point & & & \\
\hline
$Z$ & 0.5 & 0.5 & 0.5 & $P_1$ & 0.627 & 0.627 & 0.246 \\
\hline
$L$ & 0.5 & 0 & 0 & $P_1'$ & 0.246 & 0.627 & 0.627 \\
\hline
$L'$ & 0 & 0 & 0.5 & $P_2$ & 0.373 & 0.373 & -0.246 \\
\hline
$L_1$ & 0 & 0 & -0.5 & $P_2'$ & -0.246 & 0.373 & 0.373 \\
\hline
\end{tabular}
\label{tab:kpoints}
\end{table}

\section{Heatmaps of spin-splitting function and symmetry-adapted functions in reciprocal space}

\begin{figure*}[ht!] 
\includegraphics[width=0.95\textwidth]{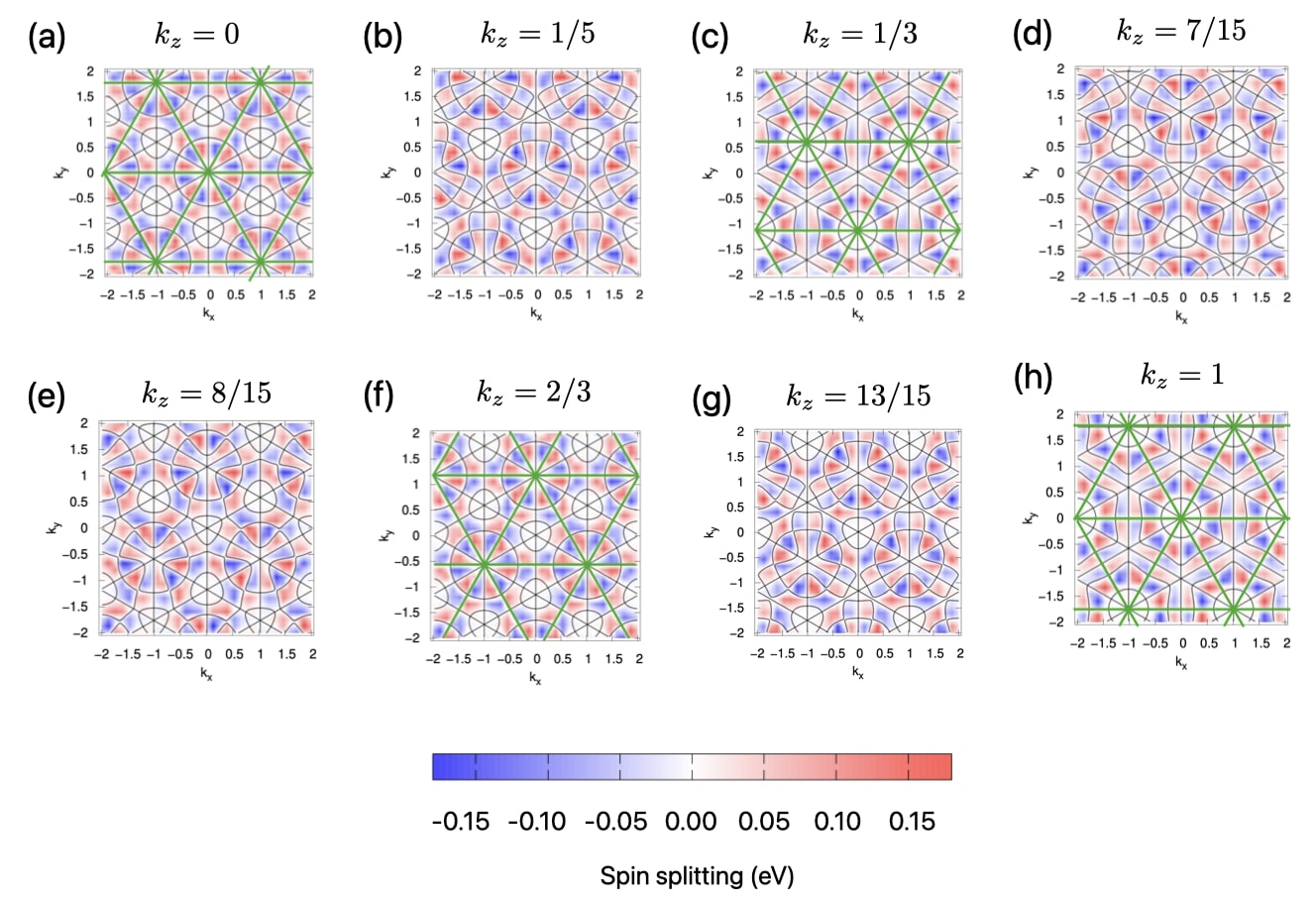}
\caption{Heatmaps of the spin-splitting function $\Delta(\mathbf{k})$, as defined in the main text, obtained from horizontal cuts of the BZ at different constant $k_z$. $k_x$ and $k_y$ coordinates are in units of $2 \pi / a$, with $a = 5.668$ \AA\ the lattice constant of BFO, while $k_z$ is given as a fraction of the $\Gamma$-$Z$ distance. Nodal elements are highlighted with black solid lines, which were generated using the 9-shell fitted function. Two-fold nodal lines are additionally highlighted in green for clarity. The nodal surfaces in the region where the function is very close to zero (white areas) are accidental nodes of the fitted function not discernible in the heatmap data.}
\label{figs1}
\end{figure*}

\begin{figure*}[ht!] 
\includegraphics[width=1.0\textwidth]{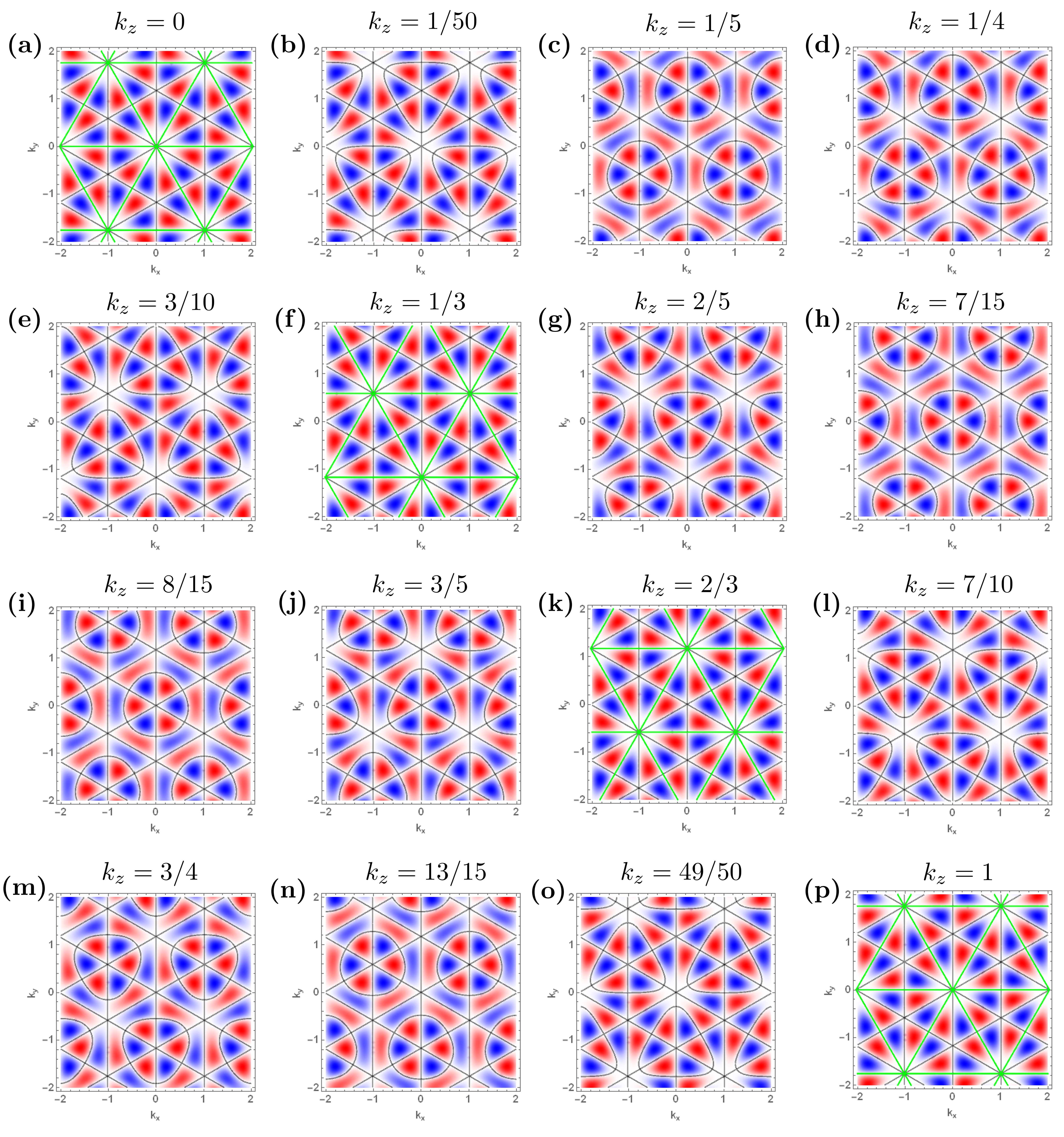}
\caption{Heatmaps at constant $k_z$ values of the symmetry-adapted combination of plane waves corresponding to the first non-trivial star in BFO, $W_1(\mathbf{k})$. The heatmaps depicted correspond to (a) $k_z=0$, (b) $k_z=1/50$, (c) $k_z=1/5$, (d) $k_z=1/4$, (e) $k_z=3/10$, (f) $k_z=1/3$, (g) $k_z=2/5$, (h) $k_z=7/15$, (i) $k_z=8/15$, (j) $k_z=3/5$, (k) $k_z=2/3$, (l) $k_z=7/10$, (m) $k_z=3/4$, (n) $k_z=13/15$, (o) $k_z=49/50$, and (p) $k_z=1$, where the $k_z$ values are reported as a fraction of the $\Gamma$-$Z$ distance. The $k_x$ and $k_y$ coordinates are in units of $2\pi/a$, with $a=5.668$ \AA\ being the lattice constant of the rhombohedral cell. Two-fold axis nodal lines are highlighted in green, while all other nodal features are highlighted with black solid lines.}
\label{figs2}
\end{figure*}

\begin{figure*}[ht!] 
\includegraphics[width=1.0\textwidth]{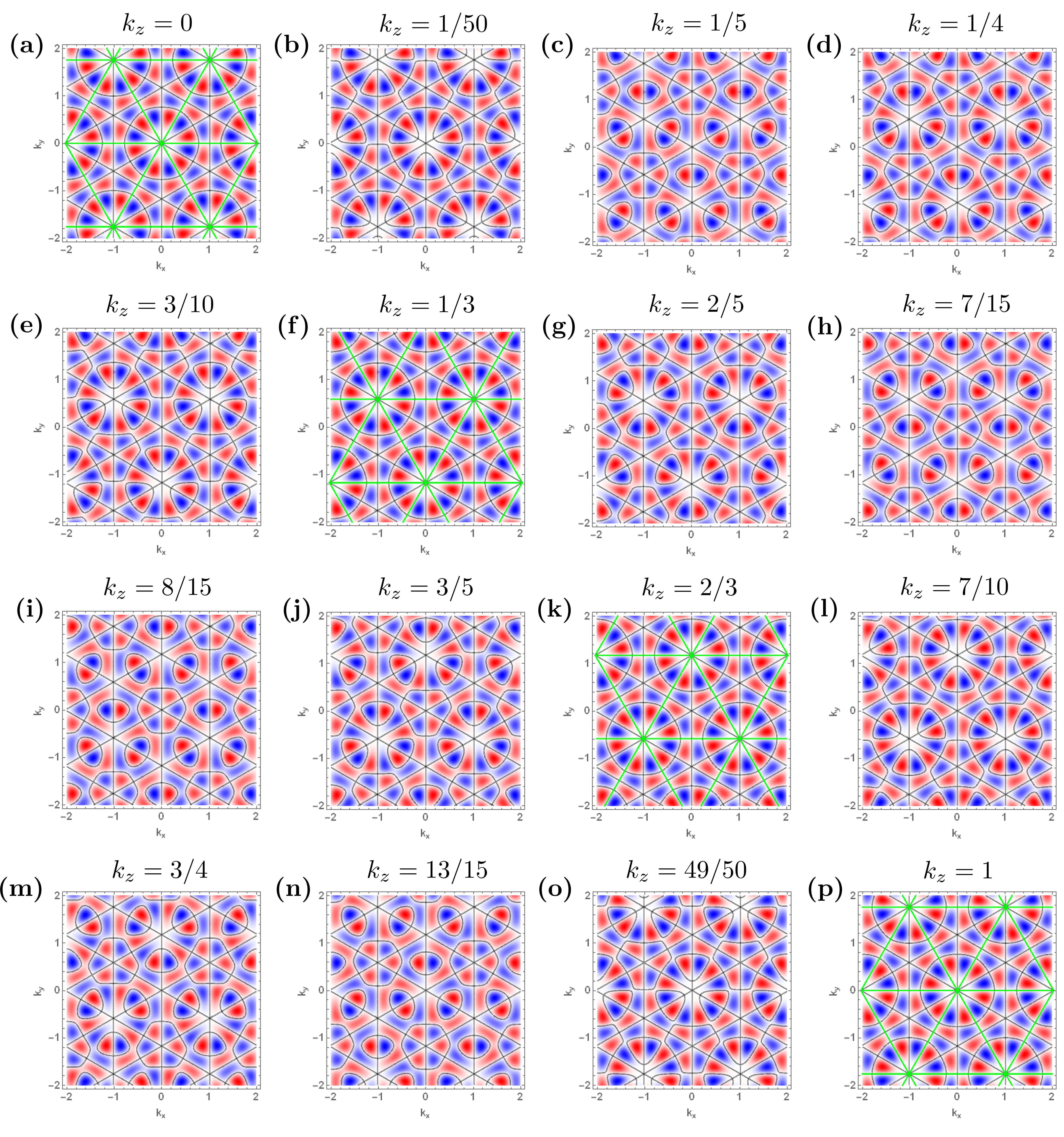}
\caption{Heatmaps at constant $k_z$ values of the symmetry-adapted combination of plane waves corresponding to the second non-trivial star in BFO, $W_2(\mathbf{k})$. The heatmaps depicted correspond to (a) $k_z=0$, (b) $k_z=1/50$, (c) $k_z=1/5$, (d) $k_z=1/4$, (e) $k_z=3/10$, (f) $k_z=1/3$, (g) $k_z=2/5$, (h) $k_z=7/15$, (i) $k_z=8/15$, (j) $k_z=3/5$, (k) $k_z=2/3$, (l) $k_z=7/10$, (m) $k_z=3/4$, (n) $k_z=13/15$, (o) $k_z=49/50$, and (p) $k_z=1$, where the $k_z$ values are reported as a fraction of the $\Gamma$-$Z$ distance. The $k_x$ and $k_y$ coordinates are in units of $2\pi/a$, with $a=5.668$ \AA\ being the lattice constant of the rhombohedral cell. Two-fold axis nodal lines are highlighted in green, while all other nodal features are highlighted with black solid lines.}
\label{figs3}
\end{figure*}

In Fig.~\ref{figs1} we provide heatmaps of the spin-splitting function $\Delta(\mathbf{k})$ for BFO, as defined in the main text, relative to horizontal cuts at constant $k_z$ in reciprocal space. On each heatmap, we highlight the contours identifying the nodes of $\Delta(\mathbf{k})$. In Figs.~\ref{figs2}-\ref{figs3} we show similar plots for the symmetry-adapted functions $W_1(\mathbf{k})$ and $W_2(\mathbf{k})$, as defined in the main text.

\begin{figure*}[ht!] 
\includegraphics[width=0.95\textwidth]{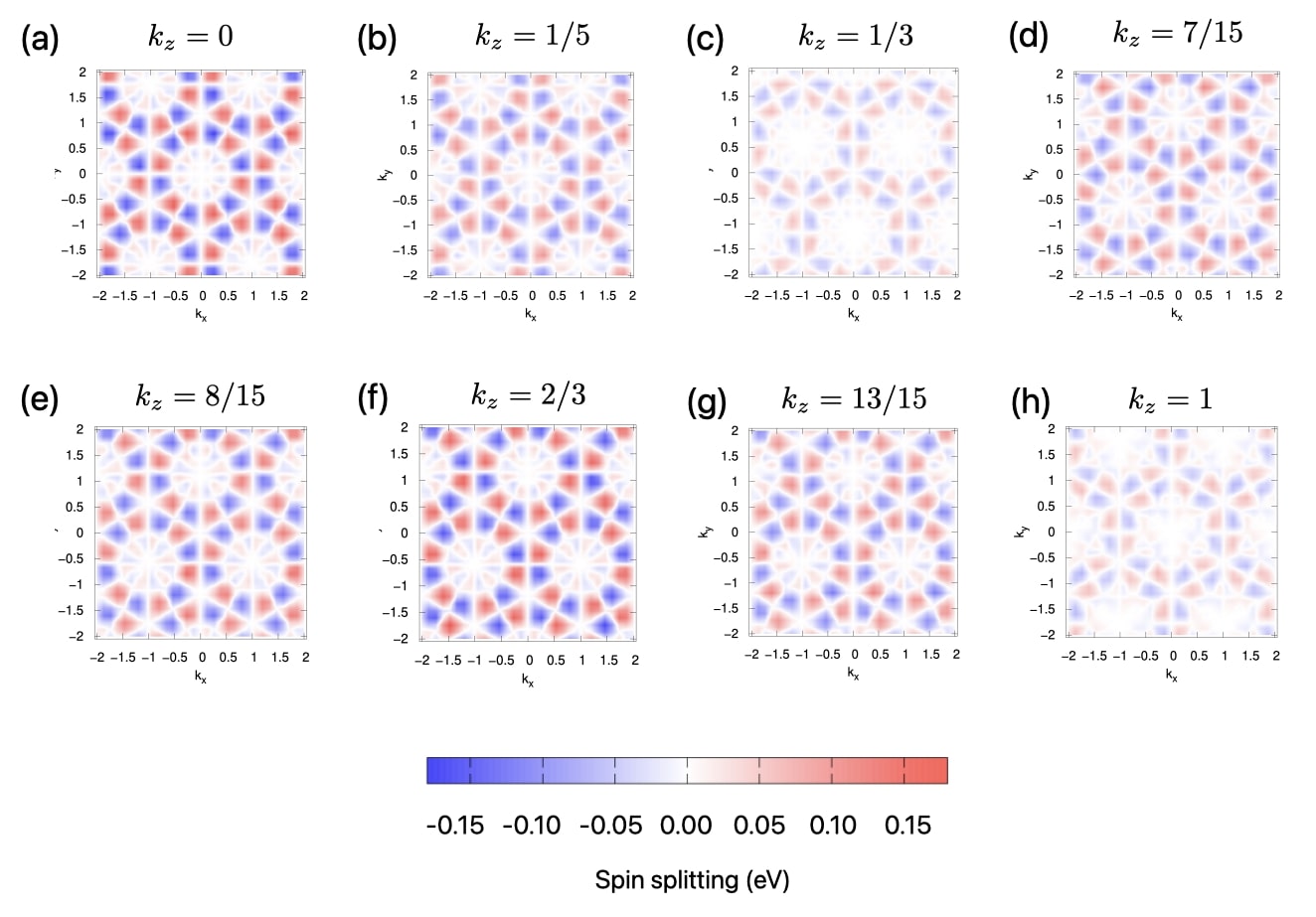}
\caption{Heatmaps at constant $k_z$ values of the spin-splitting function defined for the top two pairs of valence bands, i.e., $\Delta(\mathbf{k}) = 1/2\{\left[\epsilon_{1, \uparrow}(\mathbf{k}) - \epsilon_{1, \downarrow}(\mathbf{k}) \right] + \left[\epsilon_{2, \uparrow}(\mathbf{k}) - \epsilon_{2, \downarrow}(\mathbf{k}) \right]\}$ (see Eq.\ (1) in the main text).}
\label{figs4}
\end{figure*}

Finally, in Fig.~\ref{figs4} we provide a view of $\Delta(\mathbf{k})$ computed from the top two pairs of valence bands, i.e., $\Delta(\mathbf{k}) = 1/2\{\left[\epsilon_{1, \uparrow}(\mathbf{k}) - \epsilon_{1, \downarrow}(\mathbf{k}) \right] + \left[\epsilon_{2, \uparrow}(\mathbf{k}) - \epsilon_{2, \downarrow}(\mathbf{k}) \right]\}$. Compared to Fig. \ref{figs1}, Fig. \ref{figs4} shows fainter colors on some horizontal cuts. These correspond to regions where the two pairs of bands ($\epsilon_{1, \uparrow}(\mathbf{k}), \epsilon_{1, \downarrow}(\mathbf{k})$ and $\epsilon_{2, \uparrow}(\mathbf{k}), \epsilon_{2, \downarrow}(\mathbf{k})$) have spin splittings opposite in sign and similar in magnitude, resulting in a smaller $\Delta(\mathbf{k})$. This affects the details of the nodal surfaces, but it does not change the essential character of the continuity-enforced nodal surfaces.  

\clearpage
\newpage

\section{Coefficients of symmetry-adapted plane wave expansion of the spin splitting function for BFO}
\begin{table}[H]
\centering
\begin{tabular}{lccccccccc}
\hline
 $\text{   }\text{   }\text{   }\text{   }\text{   }\text{   }$(eV) & $\Delta_1$ & $\Delta_2$ & $\Delta_3$ & $\Delta_4$ & $\Delta_5$ & $\Delta_6$ & $\Delta_7$ & $\Delta_8$ & $\Delta_9$ \\
\hline
Least-squares & $-0.00229$  & $-0.03942$ & 0.014741 & 0.02188 & 0.02055 & $-0.00188$ & $-0.00067$ & $-0.01354$  & 0.00446 \\
Monkhorst-Pack        & $-0.00229$  & $-0.03782$  & 0.014241 & 0.02171 & 0.01753  & $-0.00105$ & $-0.00091$  & $-0.01317$ & 0.00446 \\
\hline
\end{tabular}
\caption{Values of the expansion coefficients $\Delta_s$ corresponding to nontrivial stars obtained from a least squares fitting of the spin splitting data obtained from a non-self consistent DFT calculation performed on a uniform $31\times31\times31$ $\Gamma$-centered $\textbf{k}$-mesh, and a $7\times7\times7$ Monkhorst-Pack $\textbf{k}$-mesh. The coefficients are listed in units of eV.}
\label{tab:delta_comparison}
\end{table}

\begin{table}[H]
\centering
\begin{tabular}{cc}
\hline
Star & Representative $\mathbf{R}$ \\
\hline
1 & $2\textbf{a}_1-\textbf{a}_2$ \\
2 & $2\textbf{a}_1-2\textbf{a}_2+\textbf{a}_3$ \\
3 & $2\textbf{a}_1-\textbf{a}_2+\textbf{a}_3$ \\
4 & $3\textbf{a}_1-2\textbf{a}_2-\textbf{a}_3$ \\
5 & $3\textbf{a}_1-2\textbf{a}_2$ \\
6 & $3\textbf{a}_1-\textbf{a}_2$ \\
7 & $2\textbf{a}_1+\textbf{a}_2$ \\
8 & $3\textbf{a}_1-2\textbf{a}_2+\textbf{a}_3$ \\
9 & $3\textbf{a}_1-3\textbf{a}_2+\textbf{a}_3$ \\
\hline
\end{tabular}
\caption{List of representative direct lattice vectors corresponding to the first nine nontrivial stars entering the symmetry-adapted combination of plane waves expansion for the spin splitting function $\Delta(\textbf{k})$ corresponding to the fully relaxed structure of BFO. The other vectors belonging to the star may be obtained by applying the operations in the corresponding crystallographic Laue group $\bar{3}m$ to the vectors listed in the table. $\textbf{a}_1$, $\textbf{a}_2$, and $\textbf{a}_3$ are the primitive lattice vectors.}
\label{tab:stars_representatives}
\end{table}


\bibliography{cite}